Thesis (Dissertation) for the Degree
of Master

# Mobile-Driven Incentive-Based Exercise for Blood Glucose Control in Type 2 Diabetes

by
**Wasim Abbas**

Department of Electronics, Information and Commination Engineering

Graduate School

Kangwon National University

February, 2025

Supervised by

Professor Lee Je-hoon

# Mobile-Driven Incentive-Based Exercise for Blood Glucose Control in Type 2 Diabetes

A THESIS (DISSERTATION)
Submitted to the Graduate School of
Kangwon National University in Partial
Fulfillment of the Requirements
For the Degree of

Master of Engineering

by
**Wasim Abbas**

Department of Electronics, Information and Commination Engineering

December 2024

# Mobile-Driven Incentive-Based Exercise for Blood Glucose Control in Type 2 Diabetes

Wasim Abbas

*Department of Electronic, Information and Communication Engineering*
*Graduate School, Kangwon National University*

## Abstract

    The use of mobile applications and smartphones is rapidly growing and mobile health (mHealth) solutions offer a cost-effective way to provide resources for self-management and education of chronic diseases like Type 2 Diabetes. While many applications have been developed for diabetes management, they often lack a design principle and a solid theoretical foundation or tailored to older adults, particularly those with low digital literacy. This study aims to address these gaps by developing a reward-based exercise and diabetes self-management application. This application enables users to conveniently monitor diabetes-related metrics and encourages them to manage their condition through regular exercise, medication, diet and frequent BG checks.

    We propose and create an incentive-based recommendation algorithm aimed at improving the lifestyle of diabetic patients. This algorithm is integrated into a real-world mobile application to provide personalized health recommendations. Initially, users enter data such as step count, calorie intake, gender, age, weight, height and blood glucose levels. When the data is preprocessed, the app identifies the personalized health and glucose management goals. The recommendation engine suggests exercise routines and dietary adjustments based on these goals. As users achieve their goals and follow these recommendations, they receive incentives, encouraging adherence and promoting positive health outcomes. Furthermore, the mobile

application allows users to monitor their progress through descriptive analytics, which displays their daily activities and health metrics in graphical form.

To evaluate the proposed methodology, the study was conducted with 10 participants, with type 2 diabetes for three weeks. The participants were recruited through advertisements and health expert references. The application was installed on the patient's phone to use it for three weeks. The expert was also a part of this study by monitoring the patient's health record. To assess the algorithm's performance, we computed efficiency and proficiency. As a result, the algorithm showed proficiency and efficiency scores of 90% and 92%, respectively. Similarly, we computed user experience with application in terms of attractiveness, hedonic and pragmatic quality, involving 35 people in the study. As a result, it indicated an overall positive user response. The findings show a clear positive correlation between exercise and rewards, with noticeable improvements observed in user outcomes after exercise. The incentive-based system improves diabetes self-management outcomes by effectively motivating users to participate in regular physical activity.

☐ Key Words

# CONTENTS

# Table of Contents

# Table of Figures

# I.   Introduction

## 1.1   Background

Diabetes mellitus (DM) is a major global health concern, especially type 2 diabetes (T2D), a chronic metabolic disease characterized by persistent hyperglycemia [1, 2]. Uncontrolled prolonged elevated blood glucose can lead to serious health compilation. These complications include heart disease, kidney failure, and nerve damage and vision loss [1, 2]. With the number of people affected rising from 150 million in 2000 to 537 million in 2021, the prevalence of type 2 diabetes has dramatically increased. By 2045, 783 million people are predicted to have the disease [3, 4]. This alarming trend emphasizes how urgently new approaches are needed to manage and control type 2 diabetes, particularly through lifestyle changes like exercise.

Exercise is crucial for the management of type 2 diabetes because it enhances insulin sensitivity and glycemic control. Nevertheless, many people continue to be physically inactive despite its well-known benefits. Over 1.4 billion adults globally do not meet the recommended levels of physical activity, with 32% of women and 23% of men aged 18 and older not being active, according to a World Health Organization (WHO) report [5, 6]. Effective diabetes management suffers greatly by this widespread inactivity.

Incentives-based interventions that use gamification to promote physical activity have been introduced by recent developments in mobile health (mHealth) technologies. These programs encourage people to participate in regular physical activity by providing them with external benefits, such as tangible or intangible prizes [7, 8]. Research has shown that these reward-based programs can successfully increase patient motivation, which in turn improves exercise adherence and, as a result, blood glucose control in people with type 2 diabetes [9, 10]. The incorporation of gamification into mHealth applications has great potential as a scalable and effective way to enhance T2D management outcomes by making physical activity more pleasurable and rewarding [11‑12].

## 1.2  Problem Statement

Effective management of type 2 diabetes (T2D) requires a holistic approach that integrates exercise, blood glucose monitoring, medication adherence, and dietary management to achieve optimal glycemic control and improve insulin sensitivity [13]. While exercise plays a key role, many patients struggle to maintain regular routines due to busy lives, blood glucose fluctuations, and lack of external motivation [14]. Additionally, gaps in knowledge regarding self-monitoring, medication adherence, and dietary habits further hinder effective diabetes management [15]. Mobile health (mHealth) technologies offer a promising solution by incorporating digital tools to enhance patient engagement and improve health outcomes [16, 17]. Reward-based systems integrated into mHealth applications can provide external motivation, encouraging patients to adhere to exercise, medication, diet, and blood glucose monitoring routines. By gamifying the process and offering tangible rewards for achieving individualized goals, such applications can empower patients to improve self-management behaviors [18]. Few studies, however, have explored the combined impact of incentive-based systems on blood glucose control, exercise adherence, and long-term self-management [19].

Therefore, developing an innovative mobile application that incorporates comprehensive self-management techniques, educational interventions, personalized feedback, and reward systems can address these gaps, offering an effective and practical solution for improving glycemic control and patient motivation in individuals with type 2 diabetes.

## 1.3  Research Question

This study's main goal is to investigate how exercise affects blood glucose levels in adults with type 2 diabetes (T2D), with a particular focus on how well an incentive-based system works to encourage consistent physical activity. This study will specifically investigate how incorporating an incentive-based system into a diabetes management mobile app can improve glycemic control, improve exercise adherence, and have a positive impact on overall health outcomes. To reduce participant burden and guarantee precise, real-time data collection, the study takes a user-centered approach, integrating blood glucose monitoring, medication adherence, diet and step tracking. We hypothesize that rewarding exercise and achieving goals will not only improve glycemic control but also lead to influencing participants' exercise behavior.

This study will be guided by the following research questions:

a) Question 1:

How does exercise, supported by an incentive-based mobile app, affect blood glucose levels in adults with type 2 diabetes?

- What impact does incentive-based exercise intervention have on blood glucose control, calorie burning, and overall health outcomes in individuals with type 2 diabetes?
- How do other metabolic outcomes like blood glucose, calorie intake, and calorie expenditure, as well as the user's overall diabetes management, change when a reward-based system encourages increased physical activity?

b) Question 2:

How much does incentive-based system that incorporates a medication adherence and health goals improve type 2 diabetes patients' adherence?

- How does the incentive-based system's efficacy differ depending on the baseline activity levels or various demographic groups?

c) Question 3:

What are the effects of an incentive-based exercise intervention on glycemic control, overall health, and diabetes management?

- What is the relationship between increased physical activity through an incentive-based system and long-term improvements in glycemic control?

This study seeks to fill the current gap in understanding the impact of gamified, systems on both immediate and sustained improvements in T2D management. As mobile health technologies continue to evolve, this research will provide critical insights into how personalized, incentive-driven interventions can enhance the self-management of T2D.

## 1.4　Proposed Solution

Recent developments in technology, such as mobile health (mHealth), offer new possibilities for improving type 2 diabetes (T2D) self-management through a comprehensive approach that integrates exercise, blood glucose monitoring, medication adherence, and dietary education. For instance, reward-based systems using gamification strategies can enhance patient engagement by motivating individuals to achieve personalized targets related to physical activity, blood glucose levels, medication adherence, and dietary habits [20]. In individuals with T2D, regular exercise reduces lipids, insulin resistance, blood pressure, and A1C levels [21], but many patients struggle with adherence due to lack of motivation, interest, or time commitments. By incorporating incentive-based systems into mHealth applications, users can receive tangible rewards for reaching goals such as daily steps, maintaining blood glucose within a target range, taking medication on time, and following dietary guidelines [22].

The proposed mobile application will provide a user-centered, personalized experience by integrating real-time feedback on physical activity, blood glucose levels, medication adherence, and diet. Users will receive educational prompts on self-monitoring techniques, the importance of medication adherence, and proper dietary strategies, addressing gaps in knowledge that hinder effective self-care [23]. The app will feature a goal-setting system that allows users to track their progress and earn rewards for successfully meeting targets within specified timeframes, creating a positive feedback loop that reinforces healthy habits [24]. Personalized feedback and notifications will help users stay accountable, such as reminders for blood glucose checks, exercise logs, medication schedules, and diet records.

This approach aligns with the growing body of research supporting the role of digital health interventions in improving patient engagement and glycemic control [25]. By combining reward-based incentives, educational interventions, and real-time monitoring, this app aims to provide a comprehensive and gamified solution for T2D management. The integration of exercise, blood glucose monitoring, medication adherence, and dietary education into a single system offers an innovative, scalable, and personalized strategy to enhance diabetes self-management, address motivational barriers, and improve health outcomes for people with type 2 diabetes [22-26].

Continued research will confirm the efficacy of such solutions and their potential to transform diabetes care.

## 1.5 Aims and Objective

The main goal of this study is to examine the significance of incorporating incentive-based exercise into mobile applications that can affect blood glucose in adults with type 2 diabetes. More particularly, this study aims to find how mobile applications containing gamification and reward systems can help to increase the motivation and follow-through of users with physical activity which is important for glycemic control. Focusing on the role of exercise following the incentive-based paradigm this study is to construct a detailed and patient-specific strategy for diabetes treatment.

The specific objectives of this research are:

1. To research, design and develop a mobile application that incorporates an incentive-based approach for encouraging regular physical activity in adults with type 2 diabetes.
2. To assess the effectiveness of the incentive-based mobile application in improving adherence to exercise, medication, diet and frequents BG checks routines and to evaluate its impact on blood glucose levels and other metabolic parameters in adults with type 2 diabetes.
3. To examine the relationship between increased physical activities by the reward-based approach, and long-term improvements in glycemic control in three different meal timing and overall diabetes management.

## 1.6 Scope of the Research

In this study, the research and development (R&D) of a mobile application with a reward-based system that encourages regular physical activity in adults with type 2 diabetes is investigated. This study addresses the research-based mHealth (mobile health) application development using gamification (points or virtual coins) and rewards to support and facilitate better blood glucose management.

Users will set a personalized goals and be able to see and track real-time activity such as step counting, medication record, diet information and BG levels as well as feedback on blood glucose levels. This will provide users with rewards for completing exercise goals and caloric burn goals, and encourage regular engagement in exercise. The study will also investigate the effect of the incentive motivation-based system on user adherence to exercise routines and its subsequent effect on other metabolic parameters of interest.

Usability evaluations, including surveys and user feedback, will be performed using the application to estimate how easy it is to use, and how the user feels about using the application. These assessments aim to evaluate how well the incentive-based system can encourage long-term adherence to physical activity and improve health outcomes while providing behavior changes in the self-management of diabetes. Finally, the study addresses these factors to offer a broad analysis of how gamification can enhance the use of mHealth applications to improve diabetes management and enhance the quality of life for individuals with type 2 diabetes.

## 1.7    Significance of the Study

This research's significance lies in its ability to help address key barriers to effective diabetes management through integrating incentive-based exercise programs into mobile health (mHealth) applications. However, adhering to exercise regimens is a major challenge in many people with type 2 diabetes [27], but regular physical activity is crucial for controlling blood glucose levels and preventing diabetes-related complications. Through a simple, yet effective incentive system, using 'virtual' coins or points, the study investigates how immediate rewards, in the form of tangible incentives, can affect exercise behavior and blood glucose control. The result of this study is a contribution to the emerging field of mHealth, an application of gamification to enhance engagement and development of a mobile app that rewards physical activity and thereby encourages commitment to exercise routines [28].

This research also adds to a broader understanding of how technology-driven behavior change interventions can be used for chronic disease management. The study focuses on personalized rewards and how behavior reinforcement can not only improve blood glucose control but metabolic health as well [29, 30, and 31].

Moreover, the study also fills a gap in the literature about the efficacy of incentive-based mHealth applications for both clinic outcomes and users' satisfaction. The results could play a role in future approaches to developing more holistic solutions involving physical activity tracking, personalized feedback and sustained user engagement with digital health.

## 1.8 Thesis Structure

The thesis is structured as follows: The first chapter discusses the problem statement, research questions, and proposed solutions. Chapter two presents a broad literature review of existing theories and research on using mobile health (mHealth) technology, game design, and the effectiveness of incentive-based interventions on physical activity and diabetes. Chapter three outlines the research methodology, the study design, the data collection method and analytical techniques, and demonstrates the functionality of the mobile app and reward system through diagrams. These experimental results and study evaluation are presented in chapter four, which discusses the effectiveness of the incentive-based application in improving exercise adherence while positively affecting blood glucose levels and other health metrics. The findings are discussed in chapter five, which also discusses the limitations and compares the results with previous studies. Chapter six concludes the thesis by summarizing the main results, discussing their implication for diabetes management, and suggesting future research directions on the use of rewards in diabetes management.

# II. Related Work

Diabetes self-management (DSM) mobile applications are becoming more common than ever, allowing the user to get essential features to help them manage their condition. Typically, these applications offer such functionalities as blood glucose tracking, medication reminding, diet guidance and exercise tracking [32]. Mobile technology has also been shown to positively impact DSM and glycemic control [33] and such research suggests a potential for such applications to improve health outcomes and improve patient engagement [34]. Yet, there is a demand for mobile applications that serve the needs and challenges of people in a different way than what is being done by others. The specific needs of this demographic are the focus of this literature review, which will comprehensively explore existing research in this area, focusing on studies that specifically address these needs.

## 1. Existing Diabetes Self-Management Applications

Many studies have been conducted on mobile applications intended to support better diabetes self-management, but each has unique insights and recommendations. For instance, a diabetes app developed by Ehsan et al. [35] following health guidelines received positive usability evaluations from a variety of stakeholders, including patients, IT specialists, and medical professionals. However, the study highlighted a crucial oversight in the design process: This highlights the need to involve end users (diabetic patients) in the app development.

Similarly, the iOS app DiaFit [36] aimed to address the importance of improving proactive self-management techniques and patient-provider communication concerning obesity and type 2 diabetes. Despite DiaFit's good intentions, the gaps between research and practical medical solutions were revealed by scheduling delays and the lack of real-world testing in primary care settings.

Besides, Fico et al. [37] introduced an ontology-based system offering personalized treatment advice for diabetes patients, for instance, meal suggestions, exercise recommendations and reminders. The system demonstrated potential for alerts for timely blood sugar level management but was weak in recommendations for meals and for recognition of food and lacked important features like activity tracking and interactive visual interfaces.

Additional research has examined diabetes management applications from a wider angle. For instance, a systematic review conducted by Huang et al. [38] explored the effectiveness of using mobile health interventions on diabetes self-management and found improvements in adherence to lifestyle changes and glycemic control. However, it also noted that user retention and sustained engagement had been a challenge across underserved populations. These findings are in alignment with previous studies that warn about the necessity of including culturally appropriate, patient-centered designs in diabetes apps [39].

## 2. Type 2 Diabetes and Physical Activity

There's no debate that physical activity plays a central role in managing Type 2 diabetes and as such, there are a lot of studies that have looked at physical activity and glycemic control. An example of such a study within the context of diabetes is that of Sigal et al. [40] who conducted a randomized controlled trial evaluating the effect of aerobic exercise, resistance training or a combination of both on blood glucose levels in subjects with Type 2 diabetes. Despite the improvements, the study found that both aerobic and resistance training were effective at lowering HbA1c levels for those with diabetes, but combined they had a significantly greater reduction than either exercise alone, implying the additive benefits of having diverse physical activity. Nevertheless, the results of this study suggest that research for personalized exercise recommendations given real-time blood glucose monitoring is lacking.

In another study, Balducci et al. [41] also investigated the effect of structured exercise programs on insulin sensitivity and cardiovascular risk factors in Type 2 diabetes patients. The exercise improved both insulin sensitivity and cardiovascular health and created an impact with the research showing improvements in insulin sensitivity and cardiovascular health. However, the study also pointed to problems of long-term adherence to these exercise regimens, essential to maintain the health impact of physical activity in the treatment of type 2 diabetes. These results indicate a lack of research on how to stimulate sustained engagement in physical activity over the long term.

Additionally, the work of Colberg et al. [42] further supported that physical activity should be combined with proper nutritional strategies, to reduce glycemic control in people with diabetes. Although their research showed that even moderate amounts of physical activity are associated with regulated postprandial blood glucose, challenges exist to render exercise programs at a level that is personalized and fits well into patients' daily routines. This underlines the fact that personalized exercise programs based on individual needs and constraints, especially in older adults and persons with physical limitations, are not well developed during the development stage.

## 3. Mobile Application Targeting Physical Activity in Diabetes Management

Several mobile apps have been developed for exercise in diabetes management, some with varying degrees of success. For example, Wu, Y. et al. [43] conducted a controlled study to test a mobile app designed to facilitate physical activity and dietary tracking with diabetes patients to determine if they can improve glycemic control. For users of the app, the results showed significant reductions in fasting blood glucose levels. However, the study also found that user engagement tailed off sharply after the 1st month, a common problem for diabetes apps in maintaining long-term user engagement.

In a similar study, Fukuoka and his co-authors [43] evaluated the use of a mobile app aimed at increasing physical activity in sedentary people with Type 2 diabetes. Although the app worked to increase daily step counts during the first few weeks of use, the study concluded users' motivation faded over time as it could not sustain such gains. This points to a crucial gap in diabetes app design: the absence of a strong, long-term motivational framework that would keep a user involved in exercise after the phase of the initial involvement.

Goyal and Cafazzo [44] also reviewed mobile health technologies aimed at helping patients with diabetes make lifestyle changes. Their review found that these apps may increase activity levels in the short term, but long-term success depends on features of self-determination, self-efficacy and continuous feedback features that are lacking in many of the apps available today.

## 4. Gamification and Incentive-Based Interventions in Diabetes Management

The use of gamification and incentive-based interventions as tools to facilitate behavior change in chronic disease care, especially diabetes care, has become popular in recent years. Using game design elements, such as points, badges, and leaderboards, these interventions create interventions aimed at increasing user engagement and adherence to healthcare behaviors. Furthermore,

AlMarshedi et al. [45] carried out a study using the "Diabetes Game" app that rewarded the users' activity when they recorded their blood glucose intake, exercise and diet. The app also resulted in significant improvements in blood glucose control compared to participants in a control group. However, the paper found that while gamification has an initial effect on motivation, the engagement levels started waning over time, which suggests more sustained incentive mechanisms are required if one were to retain long-term adherence.

In line with that, Pagliari et al. [46] assessed whether a points-based system of rewards for exercise and healthy eating can be used in a diabetes mobile app. The study found that although the app was successful in increasing short-term physical activity, it didn't motivate users to change their lifestyles; users soon started to cheat the system by choosing less strenuous exercises or inflating their activity levels to receive rewards. This demonstrates how crucial it is for apps to confirm that user-reported data is accurate and to include more significant rewards that would encourage genuine, long-lasting behavioral change.

Kerner and Goodyear [47] also point out that financial rewards may very well serve as an incentive for physical activity among Type 2 diabetic patients. Participants in that study were given money to achieve certain exercise goals. The incentive boosted physical activity during the trial but long-term follow-up showed that participation stopped when the financial incentives were withdrawn, and activity levels dropped back to baseline. These results imply that extrinsic rewards alone are unlikely to engender lasting behavior changes.

To overcome this difficulty some researchers have explored the principles of Self-Determination Theory (SDT) combined with extrinsic rewards. To implement SDT principles in a diabetes management mobile health app, Mitchell et al. [48] gave users monetary incentives in addition to tailored feedback that prioritized independence, competence, and relationships. When extrinsic rewards and SDT-based interventions were paired, user satisfaction increased, engagement was longer in duration, and absent extrinsic rewards, users remained engaged longer. This illustrates how SDT and incentive-based motivation can be joined to lead to longer-lasting behavior change.

However, very few studies have successfully balanced intrinsic motivation and extrinsic rewards to enable diabetes management applications. For the majority, it's all about outside incentives: money or points, not the means to deeply engage users in a psychological sense, i.e. allow them more control over their health choices. The existence of this gap offers a rationale for future research to explore how both the reward systems as well as SDT-based intrinsic motivational frameworks can be more effectively integrated into app design.

# III. Proposed Methodology

Figure 1 shows a comprehensive data workflow and recommendation model for an incentive-based mobile application to manage health and exercise routine, especially for diabetes management. The input section highlights three key components: including user registration or login by various methods (manual, Google account, Kakao account), user profile (such as name, age, gender, height, weight, target blood glucose levels, and exercise status), and integration of three healthcare sensors (USDA for food calories, Google Fit for exercise and step count, and a glucose sensor for blood sugar monitoring). After receiving the input, data passes through several stages of processing. The data acquisition service collects the input information, and sensor data and feeds it to the data processing stage for analysis. After that, the application continues to goal identification, where personal health goals related to exercise and blood glucose management are created. Next, based on these goals, a recommendation engine recommends particular actions the user can take, for instance, exercise routines or dietary adjustments. Finally, the application then employs a reward point's distribution mechanism to encourage user behavior and promote adherence to the suggested actions. This data flows back into a central data repository so it can be stored and used later. Through an android smartphone interface, users see their results and track their health progress in an engaging and data-driven format. Each of these phases is comprehensively presented below:

## 1. Application Input

The system's input phase initially starts when the user signs up or logs in to the application. Users can choose to log in manually or use accounts from third parties like google or kakao. Upon registering, users must enter personal information like name, gender, age, height, weight, and exercise status in addition to specific health data like average daily caloric intake and target blood glucose levels. The creation of customized health recommendations is based on this individualized data.

In addition to user-provided data, the application integrates databases and health sensor data. These include the USDA food database for tracking calorie intake, google fit for tracking physical activity (e.g., step counts), and bluetooth-enabled glucose monitors for real-time blood glucose readings. Data from these sensors is gathered by the client (smartphone), which securely transmits it to the server via REST APIs.

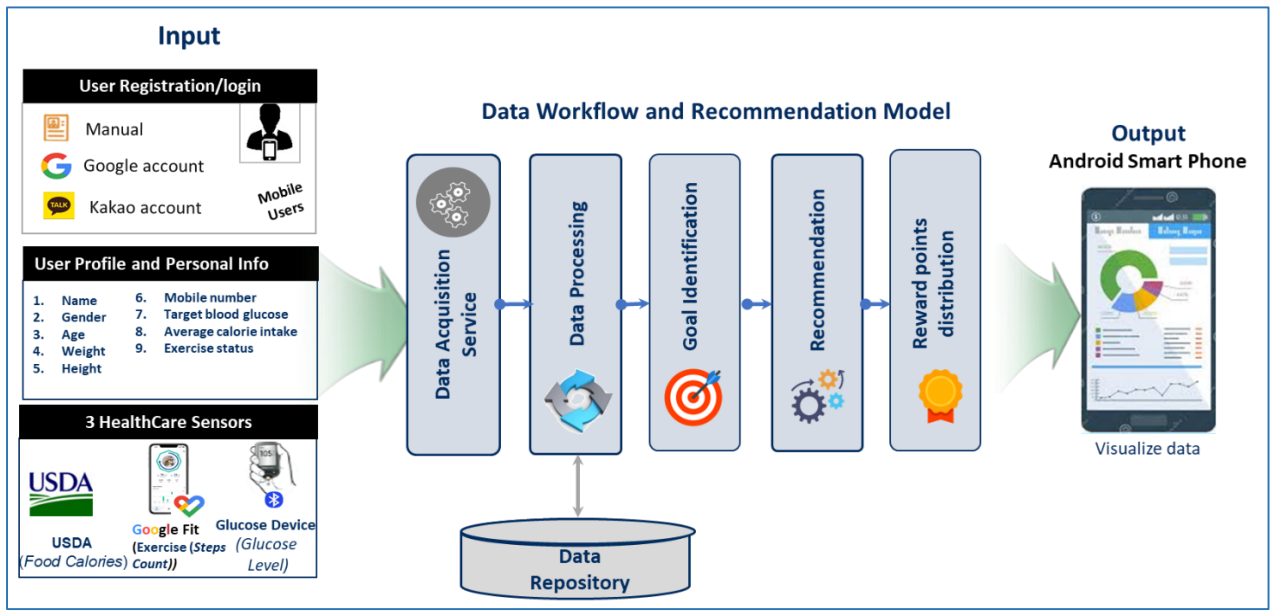

Figure 1. Proposed Mobile Application Architecture

### 1) USDA Database for Food Calories

A comprehensive source of nutritional data, including calorie counts for thousands of food items, is the USDA (United States Department of Agriculture) food database. The USDA created it to offer a standardized reference for food calories, nutrients, and other dietary components. Meal planning and nutrition tracking are two common uses for this database. Through the use of an API (Application Programming Interface), the USDA database is integrated into the application, enabling it to obtain calorie data when users record their food consumption. This aids in the app's precise calorie consumption calculation and dietary suggestions based on the user's desired level of health.

### 2) Google Fit

Google Fit is a platform for tracking health. It uses sensors or connected wearables to track a variety of physical activities, including heart rate, exercise duration, calorie burn and step count. Google fit was created to encourage users to lead healthy lives by tracking their daily exercise routines and health goals. The app's integration with google fit enables automatic syncing of activity and exercise data. Google's APIs are used to retrieve this data, giving the app real-time access to users' steps, workouts, and other physical activity metrics. This allows the app to determine how many calories are burned and monitor the user's progress toward their fitness objectives.

### 3) Bluetooth-Enabled Glucose Monitor

Devices that measure blood glucose in real time include glucose monitors with bluetooth capabilities. These monitors, also known as glucometers, enable smooth data transfer by connecting wirelessly via bluetooth to smartphones or other devices. The application incorporates these monitors to automatically record users' blood glucose levels throughout the day. Tracking glucose trends and providing individualized dietary and exercise recommendations depend heavily on this data. The application uses bluetooth connections to retrieve this data, saving users from having to manually enter data and giving them real-time insights into their blood sugar levels.

## 2. Data Workflow and Recommendation Model

The data workflow and recommendation model start with, the data acquisition service gathering information from the user and connected health sensors. After that, this data is transferred to the server, where it is processed and stored safely. The server uses SSL protocols and other security features offered by the paid hosting service for increased protection. When data is received, the server analyzes it in real-time to determine the user's health, focusing on blood sugar levels, exercise routines, and food consumption.

Following the data processing phase, the system performs goal identification, where personalized health and fitness targets are created based on the user's current status. The system dynamically adjusts these goals based on real-time data such as glucose levels and exercise performance. Using this data, the app creates practical health recommendations that are intended to increase the user's physical activity and help them better adhere to their blood glucose goals.

The recommendation phase uses these objectives to provide targeted health advice, like suggesting dietary changes or customized exercise regimens. Users earn reward points as they finish these tasks or follow the advice. The distribution system for reward points functions as a gamified incentive structure, allowing users to accumulate points for reaching fitness goals. This component is essential for sustaining user motivation, especially for those who are coping with long-term illnesses like Type 2 diabetes.

### 1) Data Acquisition Service

The first step of the workflow is data acquisition, which includes obtaining information from various sources. This includes data that the user has provided, like age, height, and weight, as well as health indicators like blood sugar, caloric intake, and exercise history. In addition, the USDA food database offers comprehensive information on food calories, and health sensors such as google fit monitor physical activity. Real-time blood glucose readings are tracked by glucose monitors with bluetooth. The client-side smartphone application collects all of this data, which is then safely sent to the server via REST APIs. The system guarantees data integrity and confidentiality through SSL encryption and other safeguards put in place by the hosting provider.

### 2) Data Processing

Data collected gets processed within the data processing stage where it's systematically cleaned, structured and organized. The raw data from multiple sources including google fit, USDA databases, and glucose sensor data is cleaned, synchronized, and ensuring data integrity as it passes through the system. This cleaned data is then divided into categories that are of interest, such as monitoring the effects of exercise on blood glucose levels at various times of the day (e.g., fasting, post-meal, and pre-meal). Once processed, the data is then stored in a centralized data repository, the online server which acts as a basis for further future analysis and goal setting. The availability of this structured data makes it possible for the system to examine trends for a user and also analyze their health condition and make personalized recommendations.

### 3) Goal Identification

The goal identification phase starts after data processing, during which the app analyzes the data to generate customized health goals. Through an analysis of the user's calorie intake, exercise routines, and glucose patterns, the system creates customized goals that cater to particular health requirements. In this case, if a user continues to have persistently high post-meal glucose levels, the app may recommend increased physical activity as a target to better regulate glucose levels. Real-time data is used to dynamically modify goals in response to the user's progress. Personalized recommendations are based on these adjusting goals, which keep the user engaged with relevant, actionable goals that promote long-term health management.

### 4) Recommendation

The system creates customized health recommendations during the recommendation phase after the goal has been identified. These suggestions are tailored to the user's specified health objectives. For example, the system might suggest a particular exercise to burn off extra calories if it determines that the user has a surplus of them. Alternatively, the system may recommend more physical activity to help lower blood glucose levels back within the desired range if the user's blood glucose level is elevated. These suggestions are regularly updated by the app using data from real time, giving the user easily actionable guidance.

### 5) Reward Points Distribution

Finally, reward point distribution is the last step in the process, where users are paid according to how well they adhere to the recommendations and achieving health goals. Users are required to set specific goals for maintaining blood glucose levels within the target range, adhering to medication schedules, achieving calorie expenditure, and engaging in regular exercise. The system incentivizes these efforts by awarding points as rewards when users meet their goals. Additionally, users are encouraged to check their blood glucose levels frequently, and if the levels fall within the target range, they receive additional rewards. This reward-based approach motivates consistent self-monitoring and adherence to healthy routines, fostering improved diabetes management. This gamification feature encourages users to continue using the app and supports them in consistently managing their health over time, particularly when coping with long-term illnesses like diabetes.

## 3. Application Output

The output application is displayed through an intuitive dashboard with dynamic charts and graphs that show the processed data. These graphic representations are aimed at helping users see their health metrics clearly and understandably, as well as trends in important metrics such as exercise, calorie intake, and blood glucose levels. The data is presented in daily, weekly and monthly views to show either short-term progress or long-term trends. For example, users can see a daily graph of their blood glucose fluctuation throughout their day in connection to their meals as well as exercise sessions. Weekly patterns might show cumulative progress in terms of calories burned, exercise time, and average blood glucose levels. For over a month, users can identify consistent patterns, enabling them to see how well they are adhering to their health goals, where improvements are needed, or where they are achieving success.

In addition to making complicated health data easier to understand, these visualizations enable users to modify their routines in real-time. For instance, users can adjust their dietary intake or increase physical activity following the recommendations if they observe a spike in their blood glucose levels following specific meals. It provides the output data to help the users stay within their target health ranges by making appropriate decisions regarding their diet, exercise and lifestyle. Most importantly, this continuous tracking is very important for users of chronic conditions like diabetes who rely on these trends to know how to effectively self-manage.

In conclusion, the architecture of the system preserves a uniform and safe flow of data from the client to the server. Security (SSL, additional protocols) protects sensitive health data, while REST APIs allow real-time communication between smartphone and server. The objective of this mobile application is to increase adherence to self-management routines and improve blood glucose management for patients with Type 2 diabetes by integrating sensors into the healthcare, a reward-based incentive system and personalized health recommendations.

## 4. Proposed Algorithm

The proposed approach for the diabetes self-management app offers a comprehensive guide for users to control their blood sugar levels and exercise while receiving incentives. The details of the system are built on top of a decision-making process with inputs of fasting, blood glucose monitoring, meals and exercise. Ultimately, the proposed algorithm is essential to enabling users to make real-time decisions about their health: when and how to exercise; what actions to take based on their blood glucose reading; and how their activities translate into rewards.

The algorithm presented in the flowchart is a structured framework designed to help individuals manage blood glucose (BG) levels, achieve health goals, and maintain a daily routine focused on BG monitoring, medication adherence, diet, and exercise. The process begins with user registration, where personal, medical, and health-related information is provided, alongside the user's consent. Users are required to set target BG goals, and if these goals are not set correctly, the system recommends a proper goal range. The algorithm then evaluates the user's knowledge in four key areas: blood glucose monitoring, medication adherence, dietary management, and exercise habits. If the user lacks understanding in any of these areas, targeted educational interventions are provided or recommended. For instance, BG self-monitoring education addresses the importance of frequent BG checks and management strategies, while users unfamiliar with proper medication adherence receive guidance on taking medicines regularly and on time. Similarly, those unaware of diabetic dietary needs are educated on diet plans, food calorie control, glycemic index (GI) values, and food exchange tables. If users do not exercise regularly, the system educates them on the benefits of physical activity, its duration, intensity, and appropriate types.

　　Once the knowledge gaps are addressed, the algorithm transitions to daily monitoring. Users log in to the system, where they are prompted to set specific goals for the day. If daily data−such as BG records, exercise logs, medication adherence, or dietary inputs−are not logged, the system sends reminders at three-day intervals. Logged data is then evaluated to check for adherence to the goals. If blood glucose records are provided, the system determines whether the levels are within the target range. If the BG is outside the range, feedback is sent to highlight the failure to achieve goals. Exercise, medication, and diet records are similarly monitored to assess overall compliance. The system also includes a reward mechanism to motivate users. If users achieve their daily goals, they are rewarded based on their performance in four areas: blood glucose levels, exercise, medication adherence, and diet monitoring. If goals are unmet, feedback is provided to encourage improvement.

　　This algorithm is designed to function as a continuous loop, fostering long-term user engagement through education, daily monitoring, personalized feedback, and rewards. By integrating behavior reinforcement and targeted interventions, the system promotes accountability and empowers users to manage their blood glucose levels effectively while adopting a healthier lifestyle.

Table 1. American Diabetes Association (ADA) guidelines for Blood Glucose levels

| **Blood Sugar Level** | **Fasting (mg/dL)** | **After Meal (mg/dL) (2 hours after meal)** |
|---|---|---|
| Low (Hypoglycemia) | Below 70 | Below 70 |
| Normal (Euglycemia) | 70 - 130 | Less 180 |
| High (Hyperglycemia) | Above 130 | Above 180 |
| Critically High | Above 180 | Above 250 |

The function of this algorithm (Figure 2) is to give real-time feedback and decision-making support to people with diabetes to facilitate consistent self-management in real-time, in a gamified way to improve motivation for exercise and blood glucose management. Moreover, the algorithm was developed around real-world scenarios, wherein all cases were tested with health data. While testing scenarios, the app generated the corresponding recommendations. The algorithm's performance and accuracy were evaluated, and the efficiency and proficiency of its responses will be discussed in the following sections.

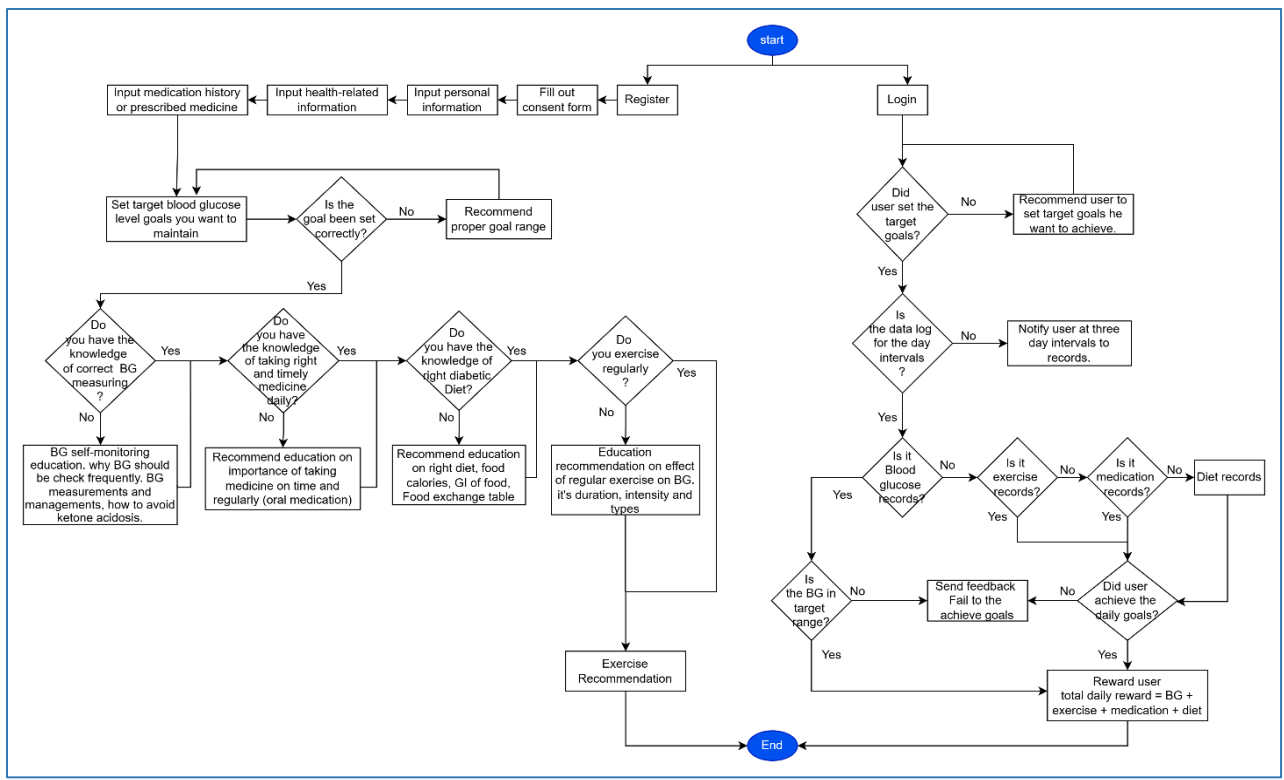

Figure 2. The DSM app algorithm was developed based on the functions and knowledge extracted in the analysis phase.

# 5. Use Case Scenario in Incentive Based Recommendation Environment

The scenarios shown in Table 2 were used to evaluate the proposed algorithm on the mobile application. This table displays the suggested exercise regimens for users based on their blood glucose levels during fasting and before and after meals. The recommendations were developed so that exercise engagement could be safely and effectively promoted while controlling blood glucose levels, which is crucial for diabetics.

The recommendation model closely interacts with the incentive-based system, in which users are rewarded according to their caloric burn during physical activity. In addition to rewarding users for engaging in physical activity, the proposed approach creates exercise points that serve as a motivating framework. Table 2 shows, how users are provided with custom feedback based on their blood glucose levels without any concern for hypoglycemia or hyperglycemia. The suggested algorithm was validated using a use case study approach, as demonstrated in Table 2, which offers a structured framework for promoting safe and efficient physical activity within the app and rewarding users. A thorough explanation of how each section works within studies is provided below.

Blood glucose (BG) levels below 70 mg/dL during fasting represent a serious health risk. To prevent further drops in blood glucose, which could lead to severe hypoglycemia, pre-exercise recommendations advise users to avoid exercise and seek medical advice. Post-exercise feedback acknowledges the decrease in blood glucose (e.g., "x mg/dL") and provides a record of calories burned (e.g., "n kcal") with reward points if exercise has already been completed. However, to avoid further risks, further physical activity is discouraged.

For users with blood glucose levels between 71‑130 mg/dL, pre-exercise recommendations encourage light-intensity exercise. The user is encouraged through reward points for completing the exercise. Completing this exercise earns users reward points based on calories burned. If after exercising, post-exercise feedback shows the drop in blood glucose, the calories burned, and the duration of exercise. Additionally, users are reminded to continue checking their blood glucose levels and to drink plenty of water.

The app encourages moderate-intensity exercise in cases where blood glucose is between 131 – 180 mg/dL. During exercise, it advises to closely monitor blood glucose levels to avoid further increase. It also encourages the activity to compete with reward points. Feedback post-exercise focuses on exercise's positive impact on blood glucose and calories burned. It is advised to stay hydrate and monitored and incentivized upon the completion of the exercise.

If blood glucose is above 180 mg/dL on fasting, it is advised not to exercise, and people are warned to check for ketones based on the risk of ketoacidosis. It recommended consultation with a healthcare professional. Post-exercise feedback emphasizes the dangerously elevated blood glucose levels and additional physical activity is strongly discouraged if the user exercises despite these warnings.

Similarly, in a Pre-meal or a Post-meal (1 – 2 hours) with blood glucose levels lower than 70 mg/dL, it is also considered dangerous. The App advised people to avoid exercise and concentrate on bringing blood glucose within a safe range. If exercise has already been done, post-exercise feedback alerts users to the possibility of hypoglycemia, gives information on calories burned and blood glucose reduction, and strongly advises stopping exercise and seeking medical help if needed.

Pre-exercise guidelines recommend exercise for blood glucose levels between 71 and 130 mg/dL. Following the incentive-based system, users are encouraged to complete the exercise to earn points based on the number of calories burned. By displaying the decrease in blood sugar and calories burned, post-exercise feedback encourages positive behavior and provides a reminder to stay hydrated and motivated to stay engaged.

Users are encouraged to participate in moderate exercise when their blood glucose levels are between 131 and 180 mg/dL. Completing the activity will earn the points. Improvements in blood glucose and calorie expenditure are highlighted in post-exercise feedback, which also suggests that they keep an eye on their blood glucose levels. While moderate exercise is recommended to help lower blood glucose levels for those with blood glucose levels between 181 and 250 mg/dL, users are cautioned against engaging in high-intense activity. Even though blood glucose is still high after exercise, post-exercise feedback indicates that exercise has improved it, and points are given upon completion. To make sure it doesn't spike any higher, ongoing observation is recommended.

Lastly, users are strongly advised to avoid exercise and check for ketones due to the risk of ketoacidosis if blood glucose levels are higher than 250 mg/dL. Consulting a doctor is advised. If the warning is ignored, post-exercise feedback indicates dangerously elevated blood glucose levels, and users are encouraged to stay away from further physical activity and consolation with doctor were recommended.

Table 2. Exercise recommendation example with each use case scenario

| Meal Type | Blood Glucose Level (mg/dl) | Recommendation and Feedback |
|---|---|---|
| Fasting | Below 70 | **Pre-exercise:** Your BG is critically low. Please avoid exercise and consult a healthcare professional.<br>**Post-exercise:** Post-exercise hypoglycemia. Please consult a doctor. Stay hydrated and avoid further exercise |
| | 71 - 130 | **Pre-exercise:** Great job! Start your exercise session with light intensity based on your calorie. You will earn points for completing the exercise.<br>**Post-exercise:** Well done! Your BG dropped to (x) mg/dL after (n) min of exercise. Keep up the good work and monitor levels, especially after meals and exercise. Burned calories: (n) kcal. Stay hydrated. |
| | 131-180 | **Pre-exercise:** (Hyperglycemia) Start moderate exercise depending on your calories but monitor exercise closely. You will earn points after completing the exercise.<br>**Post-exercise:** Well done! Your BG dropped to (x) mg/dL after (n) min of exercise. Burned calories: (n) kcal. Stay hydrated. |
| | Above 180 | **Pre-exercise:** Your BG is critically high. Please avoid exercise and check for ketones. You should consult a doctor before engaging in any physical activity due to the risk of Ketoacidosis.<br>**Post-exercise:** Your BG is still very high, avoid further exercise and check for ketones. Stay hydrated and consult a doctor. |
| Pre-Meal and Post-Meal | Below 70 | **Pre-exercise:** Your BG is critically low. Please avoid exercise until your BG returns to a safe range.<br>**Post-exercise:** Post-exercise hypoglycemia, Your BG dropped to (x) mg/dL after (n) min of exercise. Burned calories: (n) kcal. Please avoid further exercise and stay hydrated. Consult a doctor if necessary. |
| | 71 - 130 | **Pre-exercise:** You can start light to moderate exercise based on your calorie. Earn points after completing the exercise.<br>**Post-exercise**: Well done! Your BG dropped to (x) mg/dL after (n) min of exercise. Burned calories: (n) kcal. Stay hydrated. |

| | 131-180 | **Pre-exercise:** Great job! Safe to exercise. Start moderate exercise depending on your calories. You will earn points after completing the exercise.<br>**Post-exercise:** Well done! Your BG dropped to (x) mg/dL after (n) min of exercise. Keep monitoring to ensure it does not rise too much. Burned calories: (x) kcal and stay hydrated. |
|---|---|---|
| | 181-250 | **Pre-exercise:** Your BG is elevated. It is safe to exercise but avoid intense activity. Light to moderate exercise to help bring your BG downed. Earn reward for completing exercise.<br>**Post-exercise:** Good job on completing exercise! Your BG is still elevated, but exercise has helped. Continue monitoring to ensure it does not spike further. |
| | Above 250 | **Pre-exercise:** Your BG is critically high. Please avoid exercise and check for ketones You should consult a healthcare professionals before engaging in any physical activity.<br>**Post-exercise:** Your BG is still critically high. Please avoid further activity. Consumes plenty of fluid and consult a doctor. Consider adjustment to your medication and diet. |

## 6. Use Case Scenarios Evaluation

Similarly, we conducted research with ten diabetic patients using the proposed algorithm to determine the effects of weekly blood glucose checks and incentive-based exercise recommendations on the patients' health. Patients were recruited through medical experts' references and advertisements. They received guidance on using the application and were instructed to continue their prescribed medications and other health treatments throughout the study. Thus, the outcomes show how exercise and medication adherence work together. Patients were urged to log their daily blood glucose readings and perform the suggested exercises after the application was installed on their phones. To motivate engagement, a reward points system was included as a gamification feature. Over 21 days, patients used the app daily, allowing us to gather consistent data from each participant

Furthermore, we plotted the results and performed a comparative analysis before and after exercise's impact on patient health. We computed the results calculating average value by processing three weeks of data of all patients and computed results of individual users over three weeks of data comprising weekly blood glucose level and weekly based incentives before and after exercise.

This method was applied to variables such as blood glucose levels (before and after exercise), reward points, exercise duration, and the number of days to obtain the average values. The days were grouped into weeks to better analyze weekly data trends.

To calculate the results of the study blood glucose levels (before and after exercise), three types of variables are accounted for such as reward points, exercise duration, and the number of days to obtain the average values. The days were grouped into weeks to better analyze weekly data trends. We used the following approach:

1. Weekly Average for Blood Glucose (Before and After Exercise):

$$\text{Average BG (Before / After Exercise)} = \frac{\sum(BG\ measurement\ in\ a\ week)}{Numer\ of\ enrtries\ in\ a\ week} \quad \text{Equ.1}$$

2. Weekly Average for Reward Points:

$$\text{Average (Reward Points)} = \frac{\sum(Reward\ Point\ Earn\ in\ Week)}{Numer\ of\ Entried\ in\ a\ Week} \quad \text{Equ.2}$$

3. Weekly Average for Exercise Duration:

$$\text{Average (Exercise Duration)} = \frac{\sum(Exercise\ Duration\ in\ Week)}{Numer\ of\ Entries\ in\ a\ Week} \quad \text{Equ.3}$$

Where:
- $\sum$: Sum of all reading readings taken during the week.

Figure 3 demonstrates the weekly blood glucose comparison utilizing equation 1 before and after exercise. For the three-week study, a recurring pattern showed that blood glucose levels significantly decreased after exercise. In Week 1, for instance, pre-exercise levels were 160 mg/dL, and after exercise, they dropped to 148 mg/dL. This pattern continued throughout the study, indicating that exercise improves blood glucose control.

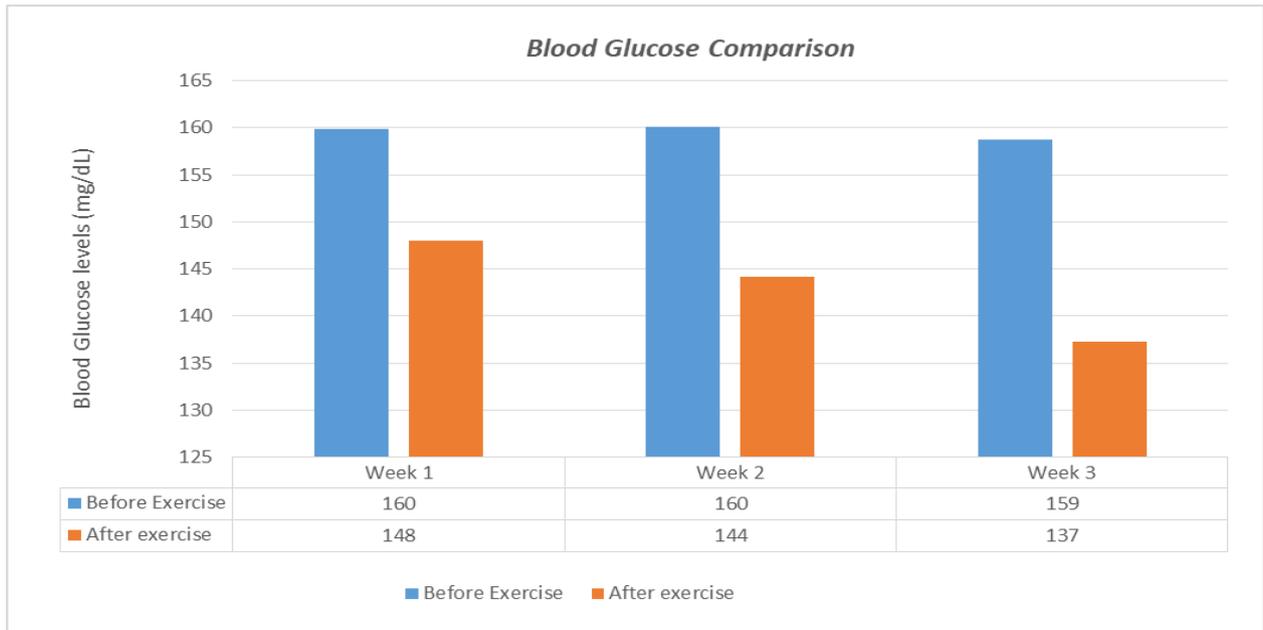

Figure 3. Comparative analysis of blood glucose before and after exercise over 3 weeks

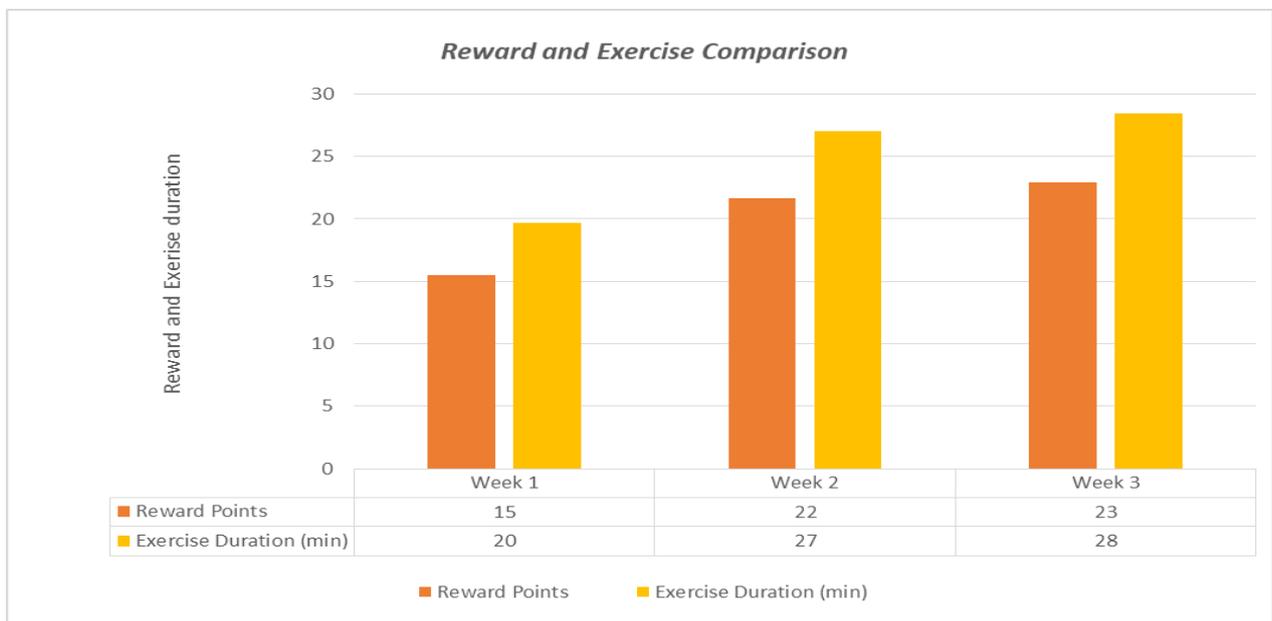

Figure 4. Comparative analysis between Reward points and Exercise duration towards Weekly improvement

Similarly, figure 4, which evaluates equations 2 and 3, further supports this correlation by illustrating the connection between reward points and exercise duration. The number of reward points earned increased as the exercise's duration increased over the weeks. This implies that users were successfully encouraged to increase their physical activity levels through the application's reward system. The information supports the hypothesis that incentives can encourage people to take up healthier habits and enhance their general well-being.

1. Average Blood Glucose (Before and After Exercise for each user):

$$\text{Average Blood Glucose} = \frac{\sum_{i=1}^{21} BG(Before/After)i}{21} \qquad \text{Equ.4}$$

2. Average Reward Points:

$$\text{Average Reward Points} = \frac{\sum_{i=1}^{21}(Reward\ Point)i}{21} \qquad \text{Equ.5}$$

3. Average Exercise Duration:

$$\text{Average Exercise Duration} = \frac{\sum_{i=1}^{21}(Exercise\ Duration)i}{21} \qquad \text{Equ.6}$$

Where:
- Summation (Σ): This adds up all the individual values for each day.
- i = 1 to 21: This represents the 21 days (or data points) collected for each user.

Figure 5 plots the mean changes in blood glucose levels for ten participants over three weeks when implementing Eq.4 to compare pre and post-exercise blood glucose levels. Through the analyzed data, it is clear that blood glucose levels decrease after exercise. For this reason, exercise is effective in managing blood glucose levels among most users. However, an exception is seen with User 3, whose glucose levels showed no change before and after exercise (156 mg/dL to 155 mg/dL), potentially indicating individual variability in response to physical activity. This lack of response may suggest that factors such as the user's overall health condition, diabetes type, or insulin sensitivity could be influencing the effect of exercise on glucose levels.

Likewise, Figure 6 shows the mean comparison of exercise duration and reward points earned from the same users for a similar period assessing Eq.5 and 6. The trend obtained is rather generalizable and it reveals that longer exercise durations result in more reward points which strongly supports the use of a reward system to encourage people to engage in physical exercises. For instance, User 9, who took the longest time doing exercises for 35 minutes, got the most reward points of 28 points. However, there were differences observed: for instance, User 6 achieved a higher number of points (25) during shorter exercise (31 min), which implies that parameters like intensity, regularity, or personal improvements might also affect the point system.

  This pattern supports the idea that encouraging exercise promotes longer periods of physical activity through evidence that as patients increase their physical activity over time, the rewards offered to them also increase. Such consistency establishes the foundation for theory regarding the application of incentive-based strategies to improve patients' overall adherence to physically encouraging exercise in the management of diabetes.

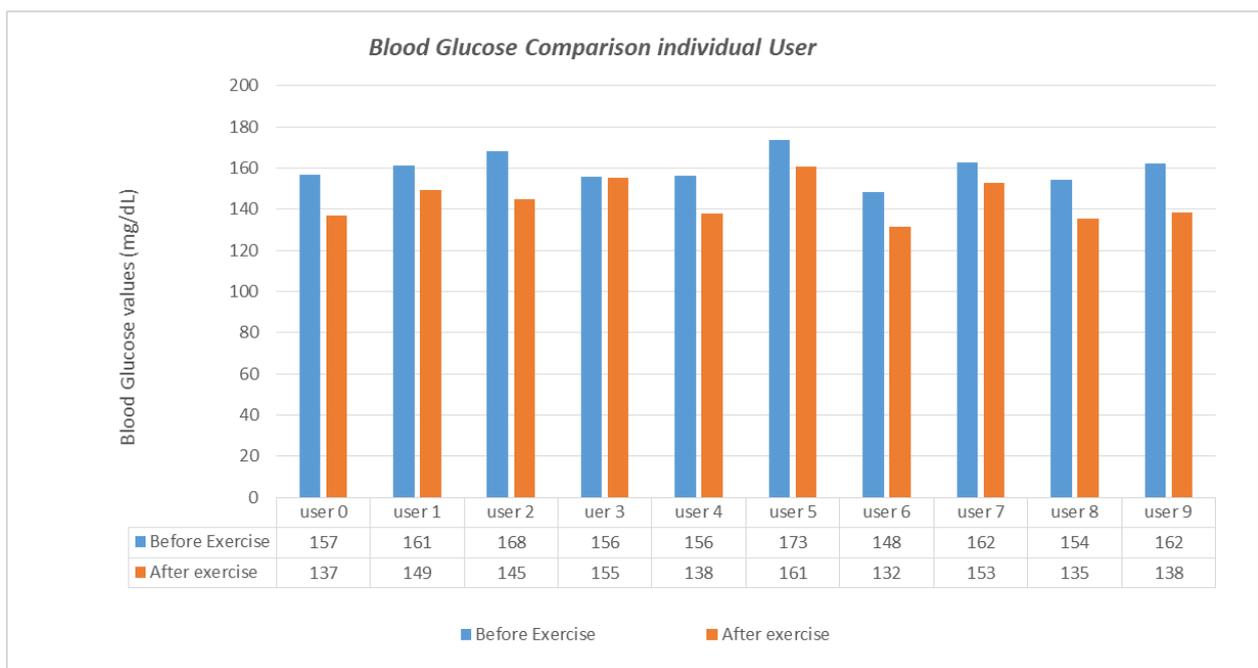

Figure 5. Patient Health performance before and after exercise towards Blood Glucose management for 3 weeks

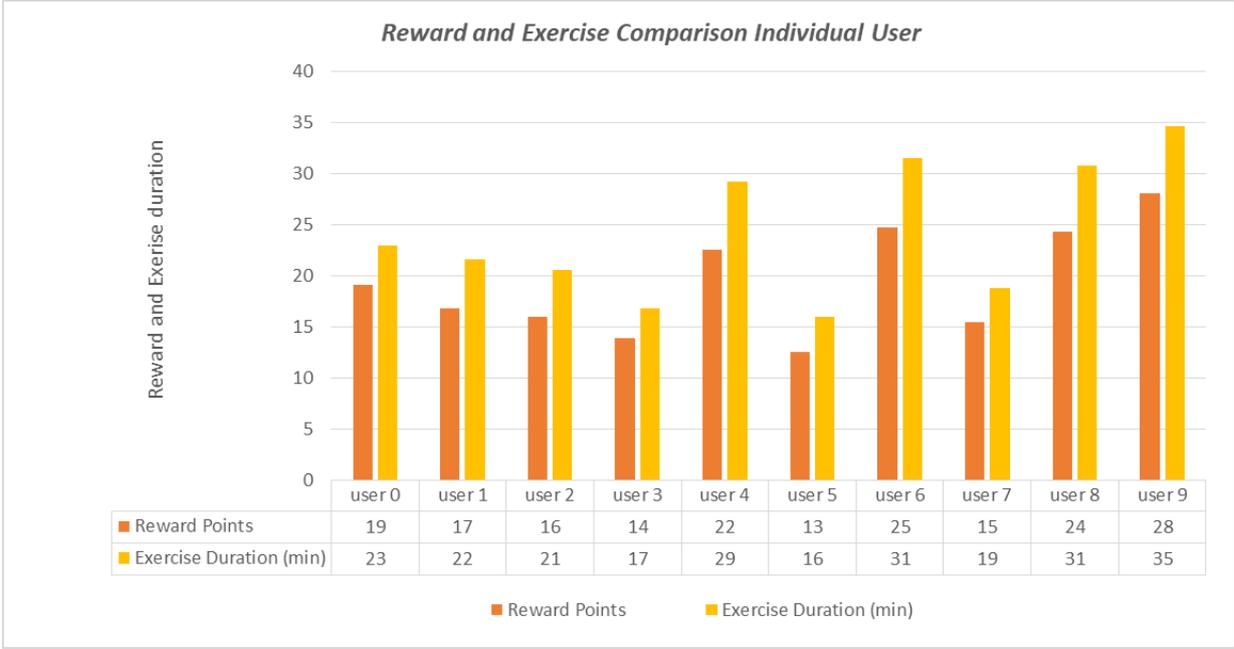

Figure 6. Comparative Analysis between Reward and Exercise impact on individual users over 3 weeks

# IV. Experimental Results, Analysis, Design and Evaluation

Figure 7 outlines the process of designing, developing, and evaluating a mobile application aimed at diabetes management to show the impact of incentive-based exercise. The development process followed a structured Software Development Life Cycle (SDLC), divided into three main phases: Analysis, Design and Implementation and Algorithm and Application Evaluation. These stages, as depicted (Figure 7), ensure that the app is developed with a clear plan, implemented based on the user's needs and validated guidelines, and continuously improved through testing and user feedback.

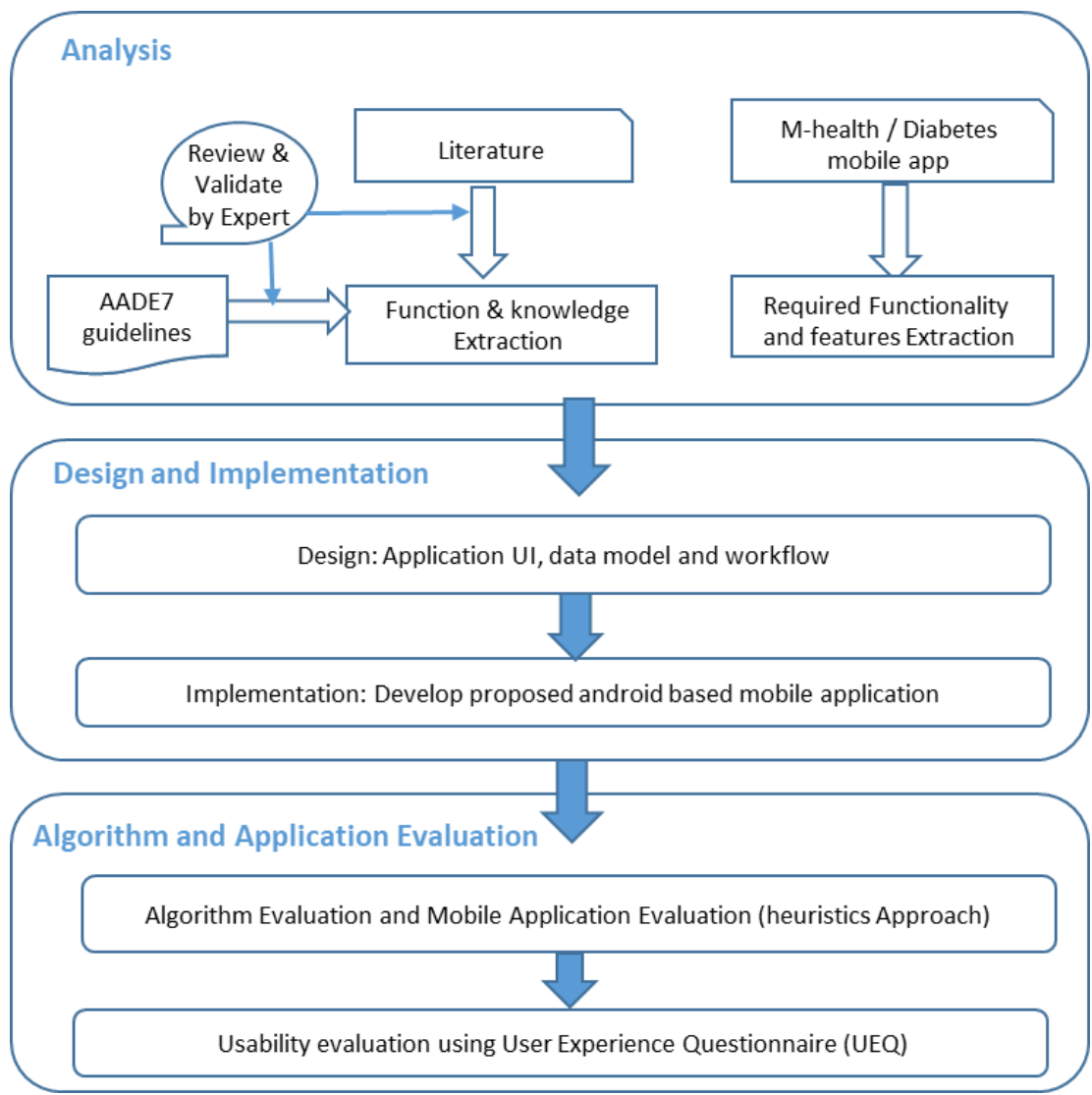

Figure 7. Three stages of Application Development Life cycle

## 1. Phase I: Analysis

The first phase begins with Analysis including the gathering of the essential requirements through a detailed review of literature, using the American Association of Diabetes Educators (AADE7) guidelines and knowing the needs and functionalities of the users through the market analysis of current diabetes management apps. Expert reviews are used to validate the process and to confirm the accuracy and appropriateness of the extracted features. The analysis is divided into two key activities: Function and Knowledge Extraction, Identifying Required Features and Functionality.

### 1) Function and Knowledge Extraction

The primary functions of the diabetes self-management (DSM) app were identified in this phase by combining a review of the literature with an analysis of user needs. DSM intervention studies, clinical practice insights, and recommendations of the American Association of Diabetes Educators (AADE7) [50] were used to derive functions. AADE7 guidelines, which emphasize self-care behaviors like healthy eating, physical activity, blood glucose monitoring, and medication adherence, were used to identify essential app functionalities. These guidelines offered a methodical way to incorporate monitoring tools, educational resources, and customized feedback mechanisms into the application. Experts then further reviewed and validated the extracted functions to ensure that they are relevant and accurate before making the app usable in real world diabetes management.

### 2) Required Features and Functionality Extraction

A methodical review of other apps available in the Google Play Store was necessary to identify the key features and capabilities of the diabetes self-management (DSM) app. We started by performing a keyword-based search to find apps that might be relevant to our study. Terms like "diabetes tracker," "blood sugar," "glucose tracker," and other similar phrases were used. A total of 194 apps were found using this search (Figure 8). However, we used a set of exclusion criteria to make sure the chosen apps matched user expectations and the unique requirements of diabetes management.

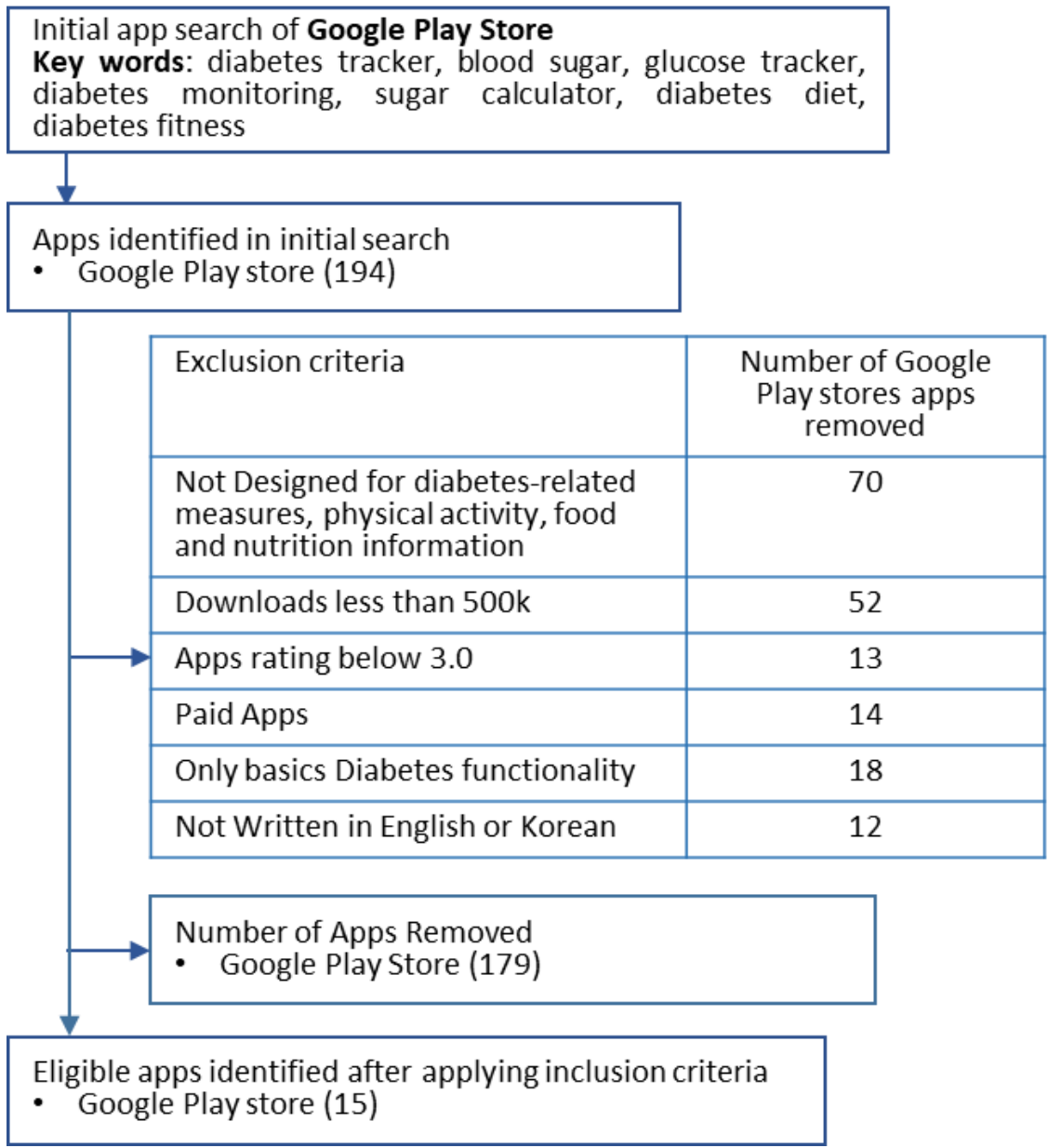

Figure 8. Analysis and selection of mHealth apps

Those excluded from the final review were apps not designed for diabetes-related measures, physical activity or food and nutrition tracking. Moreover, downloads of less than 500,000 and ratings less than 3.0 or required payment were excluded from the analysis. We also removed apps that did just offer basic diabetes functionality or weren't available in English or Korean. Applying these criteria we selected from an initial pool of apps 15 eligible applications which were further analyzed (Table 3).

Table 3. List of Diabetes Self-management Application selected for analysis

| S. No. | App Name | No. of Downloads | App Rating |
|---|---|---|---|
| 1 | mySugr - Diabetes Tracker Log | 1 Million + | 4.6 |
| 2 | Health2Sync - Diabetes Care | 1 Million + | 4.8 |
| 3 | Blood Sugar - Diabetes App | 1 Million + | 4.8 |
| 4 | Blood Glucose Tracker | 500k+ | 4.6 |
| 5 | Blood Sugar Tracker - Diabetes | 500k+ | 4.7 |
| 6 | One Drop: Better Health Today | 1 Million + | 4.2 |
| 7 | SocialDiabetes | 100k+ | 4.5 |
| 8 | Diabetes: M-Blood Sugar Diary | 500k+ | 4.5 |
| 9 | BeatO: Diabetes Care & Tracker | 1 Million + | 4.5 |
| 10 | Sugar.fit | 100k+ | 4.8 |
| 11 | Diabetes by Klimaszewski Szymon | 500k+ | 4.4 |
| 12 | Blood Sugar Diary for Diabetes | 500k+ | 4.4 |
| 13 | Blood Sugar Unit Converter | 10k + | 4.7 |
| 14 | Diabetes by Hint Solutions | 50k+ | 4.8 |
| 15 | Blood Sugar - Diabetes Tracker | 10k+ | 4.6 |

These 15 apps were downloaded, and their functionality was evaluated by entering mock data. Then we analysed things like blood glucose monitoring, exercise tracking, calorie logging, medication reminders and other tools to help with day to day management of diabetes. Features and functionalities identified during this process were compiled into a list that was then reviewed by experts in the field of diabetes care to make sure it adhered to evidence-based practices. When expert validation is completed, we then move forward with the development of our app, making sure that the app contains the most useful and necessary features identified in this analysis.

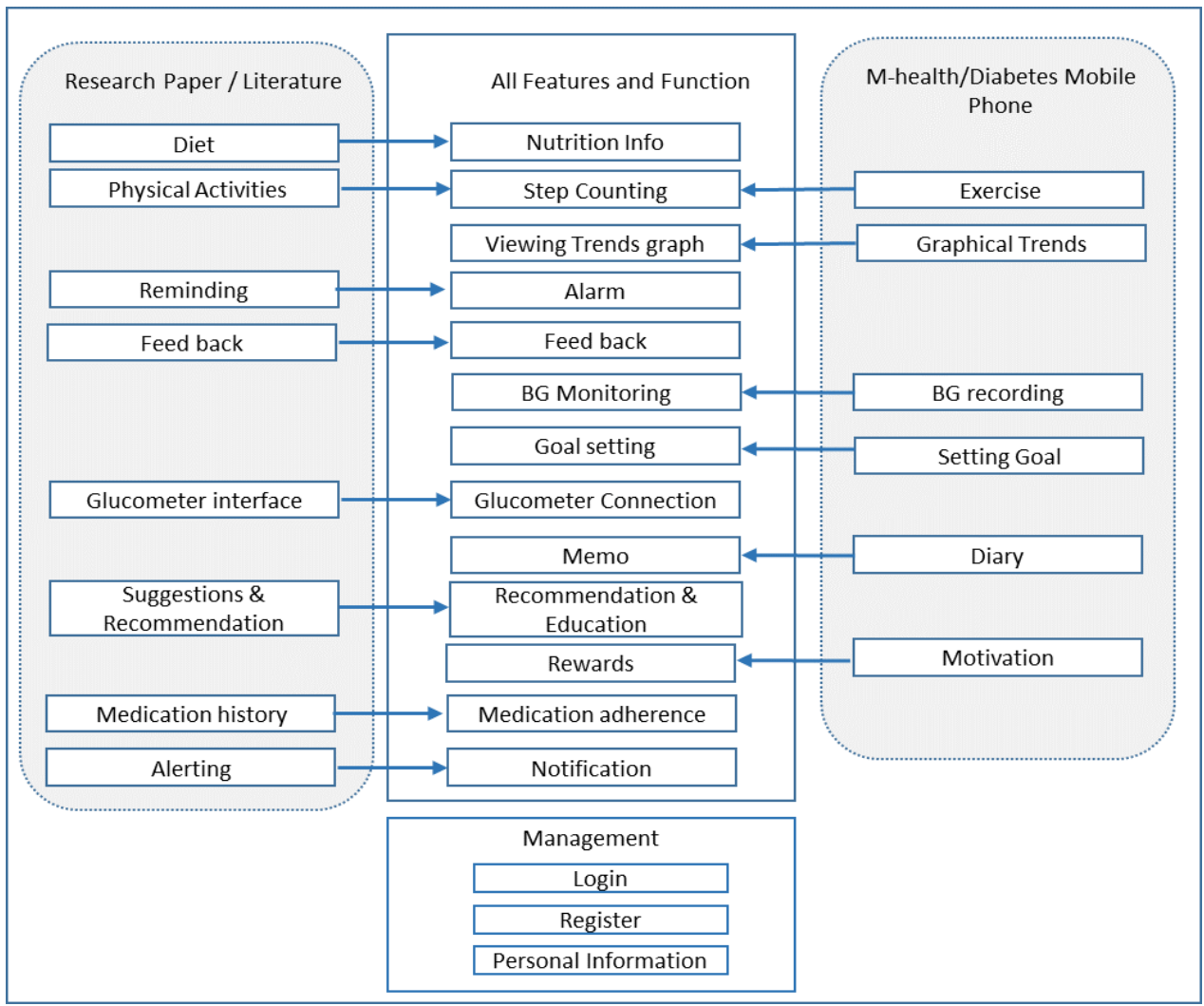

Figure 9. Sixteen main diabetes self-management extracted functions

A comprehensive set of 16 key features and functionalities to support effective self-management of diabetes was designed for the DSM mobile application (see Figure 9). An extensive analysis phase integrating the inputs from research literature and the requirements of m-health applications was performed to select these features. Essential tools for education, suggestions, feedback, and goal-setting are among the app's primary features, which help users manage their health more effectively. It contains features for monitoring diet, blood glucose levels and physical activities, and also a glucometer interface to log accurate data and monitor it.

In addition, the app also includes a gamification component that encourages users to stick to their routines by rewarding them for finishing suggested exercises. Notifications provide users with timely feedback and encouragement, while reminders and alerts help users stay on track. Tools for medication adherence, recommendations and suggestions based on user data, and alerts for regular app usage are additional essential functions. The app also has personalized management features like registration, login, and personal information management to provide a comprehensive experience. These features work together to empower the users to take charge of their diabetes management and therefore lead to a better health and well-being.

## 2. Phase II: Design and Implementation

A number of components were created during the design phase to provide a thorough overview of the system's architecture. This involved designing the user interface and prototype implementation of the proposed mobile application, as well as developing data models, a workflow, and an application user interface. Every component was carefully designed ensuring the system's consistency and functionality.

### 1) Design Application User interface (UI)

The study requirements determined during the analysis phase served as the basis for the user interface (UI) design. It was developed based on the extraction of the functions and features extracted in Phase I, with a focus on supporting efficient self-management of diabetes. Each of the sixteen identified features is designed to enhance user experience and improve health outcomes (Figure 9). Font family and colour schemes were carefully considered to create an appealing and functional design. Overall, we designed 40 user Interfaces and organized them into five major modules to make the app structure intuitive and user-friendly for diabetes management people.

### 2) Data Model

A data model, which organizes data elements and defines relationships between entities and properties, was employed to structure the application's database. In this study, we outlined all necessary data entities for the app, each associated with a specific data type (e.g., Boolean, Integer, Date, and Time), potential values, and a corresponding unit. For example, Blood Glucose (BG) was specified as an integer with the unit of mg/dl. The database we designed stores collected data about blood glucose levels, dietary intake, exercise activities, medications and personal memos. For robust and scalable features, we decided to use MySQL as our database management system from which data models and an entity-relationship diagram (Fig. 10) were generated as well. Extensive testing was done on a local server, prior to deploying the database on an online web hosting server to ensure functionality and reliability.

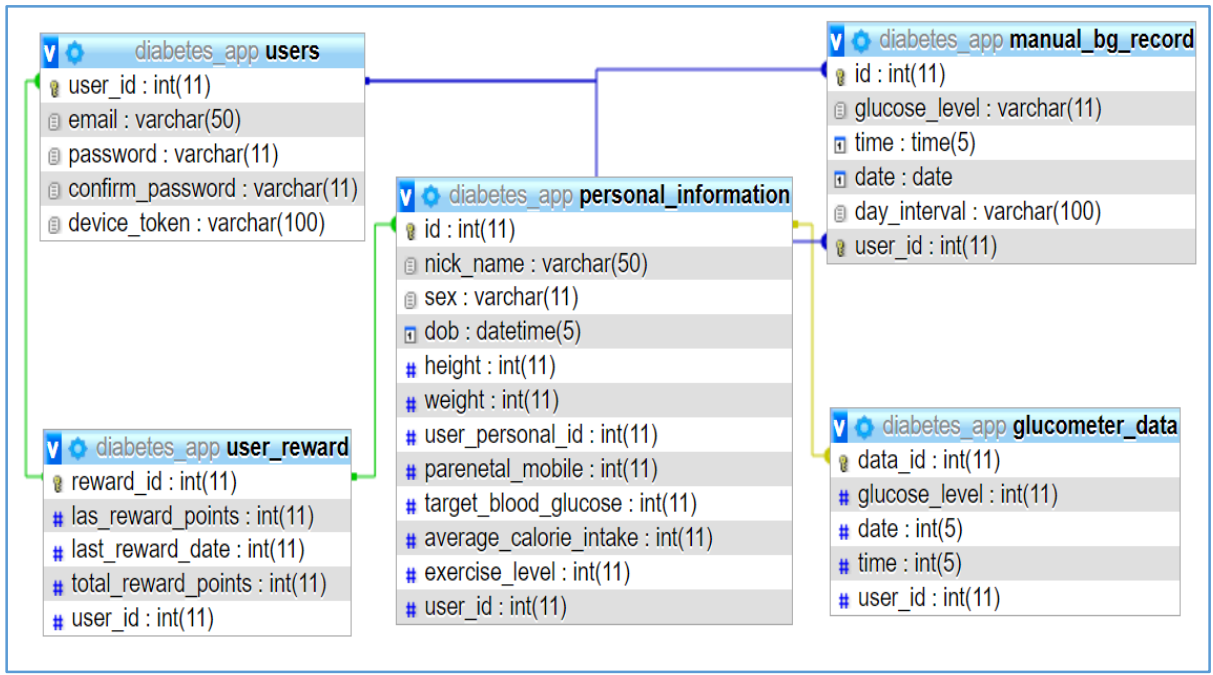

Figure 10. ERD diagram of data models

The app's data was categorized into various groups: health-related data, medication history, and goal settings, dietary records, exercise status, exercise duration, calories burned, and reward points, blood glucose readings, medication compliance, weight, height, and calorie intake. A model with potential values, data types, and units was used to represent each data category.

The database tables designed to reduce security risk by removing identifying data. Primary and foreign keys were generated by combining registration time, user ID, password, email and nickname instead of identifiable data. Data integrity as well as security are ensured with these keys. Personal information, dietary records, exercise records, rewards, blood glucose readings, and height and weight measurements, are all included in the tables.

## 3) Design Application Workflow

A total of 40 user interface screens were designed to implement five main modules along with their respective submenus. The key screens are home screen, exercise dashboard, Blood Glucose information, medication adherence tracking, and nutrition information are displayed in Figure11.

The activity diagram that highlights the key functions and user interface of the Diabetes Self-Management App is shown in Figure 11. This shows where a user can log in or sign up, navigate to the home screen, and see all the features source, including personal information input, education resources, and medicine management. There are specialized modules which allow users to track their daily steps, control their meal plans and monitor their blood glucose levels. Other than that, the app has data visualization features that give users a view of their health metrics trends as well as notifications to remind users of tasks especially taking medication or meeting the activity goals. The rewards system also encourages users to do tasks or achieve milestones and award points. The diagram overall shows that the app design encompasses everything users want to do to manage their diabetes effectively.

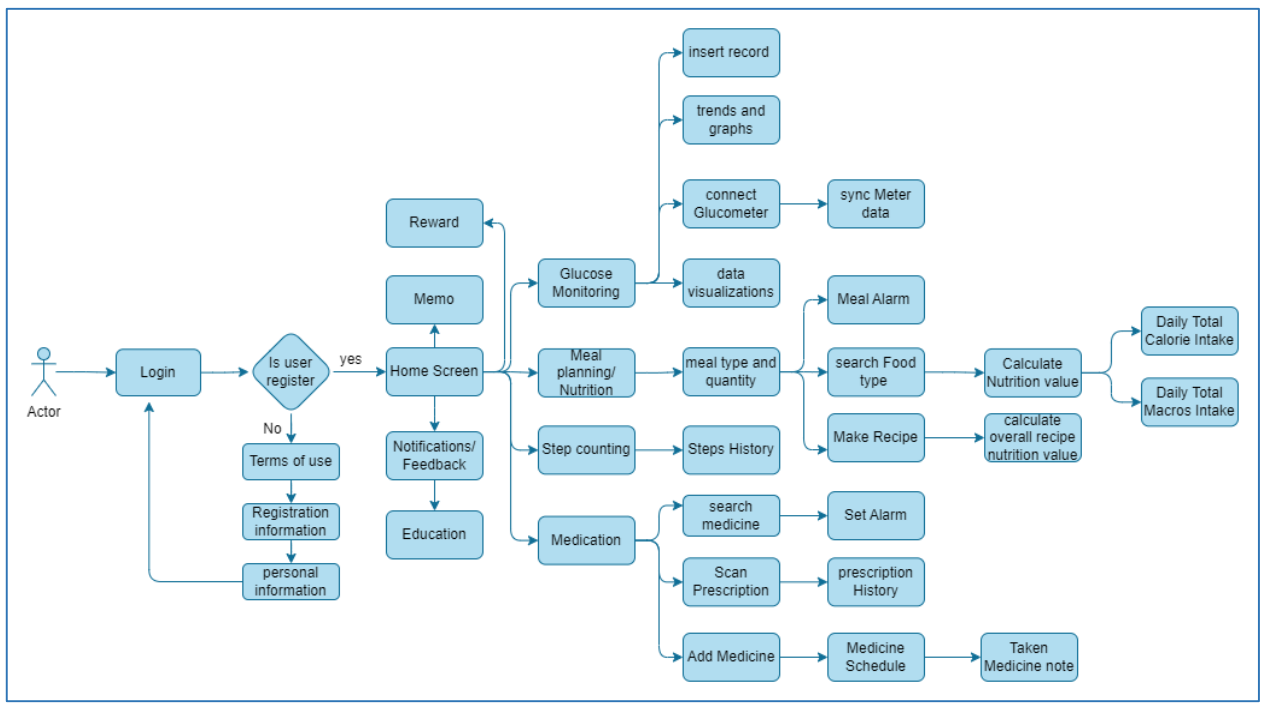

Figure 11. Proposed Application activity diagram

## 4) Application Implementation

We executed the development of the app as planned, encompassing the development of a system architecture diagram, database structure, and screen designs. Utilizing Android Studio as our primary Integrated Development Environment (IDE) and Java and Kotlin as programming languages, along with the Android Software Development Kit (SDK), we followed the standard Android development environments provided by Google. The interface was designed using XML, which allowed for different layout and screen size requirements.

We employed the Secure Socket Layer (SSL) protocol to ensure that user data was transferred securely from the application to the server. For the purpose of getting the server aspect running on Windows, we used SQLite which is well supported within the Android development platform and, so, served up from the server aspect within an Android implementation. We doubled down security measures by routing it on secure HTTPS calls for data transfer.

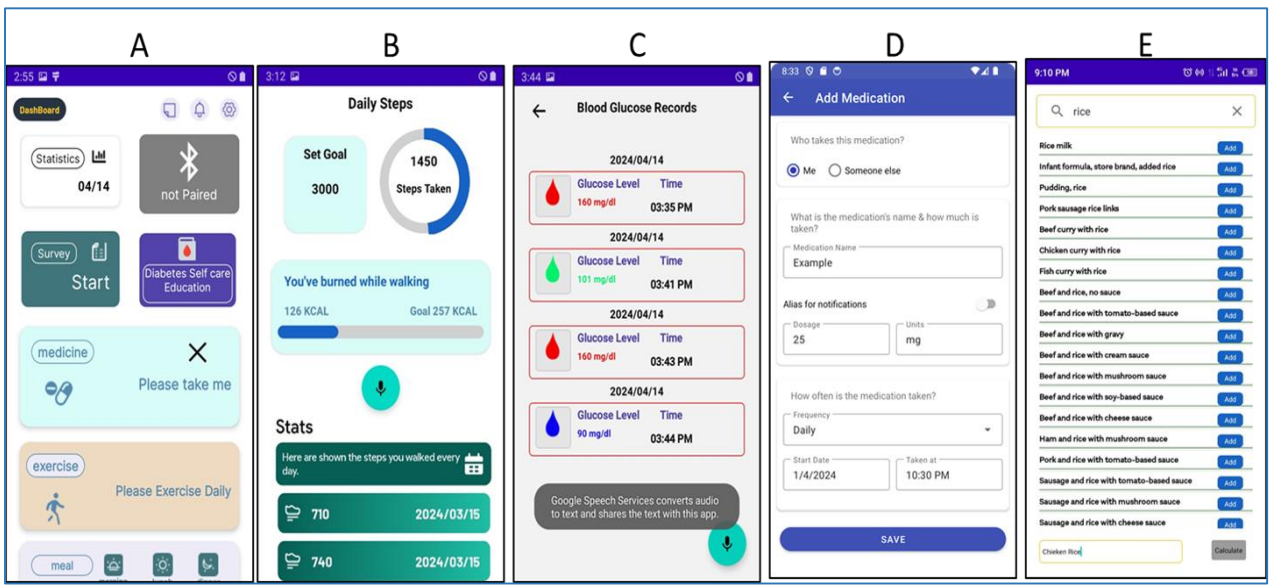

Figure 12. . (A) Home Screen, (B) Daily Steps, (C) BG recording, (D) Medication, (E) Nutrition Information

The main five modules of the app are prepared with utmost thought about how an older user would use each module; they are easy to use. We add usability guidelines for better use. The elements of these modules include home screen, blood glucose, medication, exercise, and diet (Figure 12). The functionality has been enhanced by dividing each module into multiple sub-screens. The main screen or home screen will let the user navigate to every other module very easily. It is also equipped with sub-menu options like feedback, memos and alerts. Among the features of the blood glucose module are glucose graphs, a log for monitoring blood glucose levels, data visualization, and a screen for glucometer connection. The medication module provides functions for setting alarms, scheduling medication intake, and viewing history of medications taken. The exercise module offers step counting, progress tracking, rewards, and daily step logs, while the nutrition module includes features related to meal planning and alarms for lunch and dinner reminders.

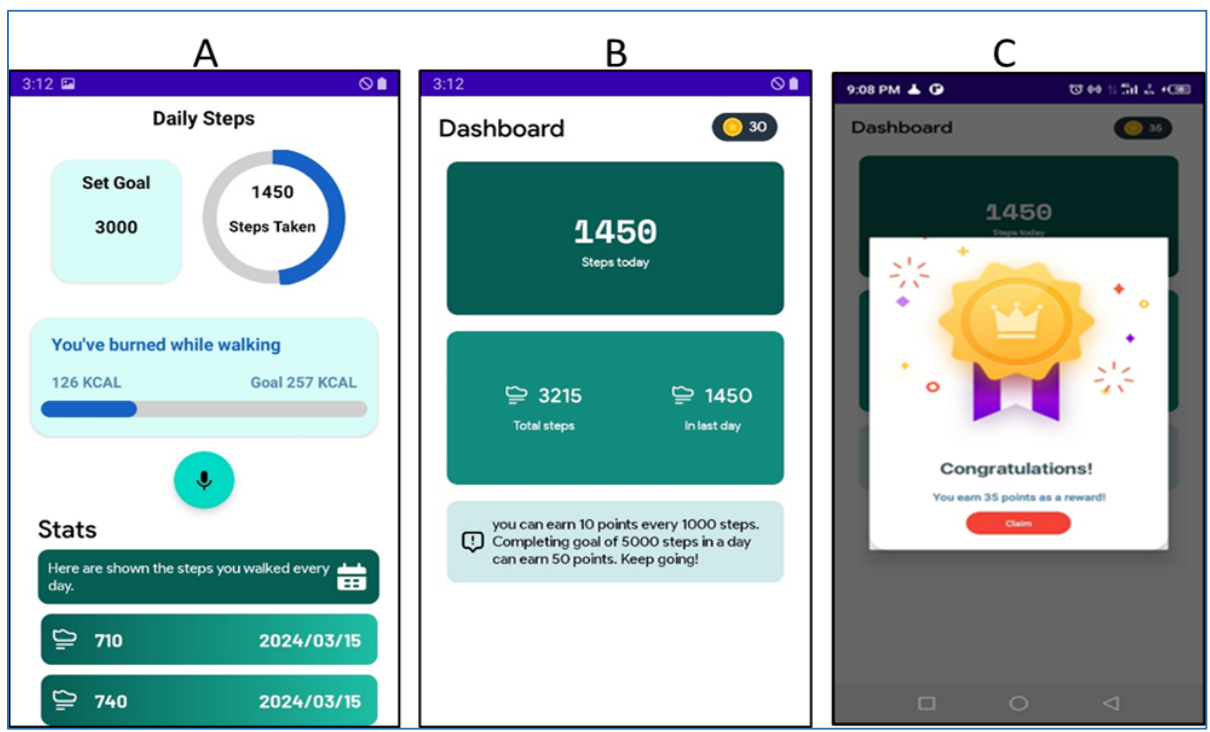

Figure 13. (A)Daily steps and goal setting, (B) Dashboard (B), and (C) Reward congratulatory dialog

The main focus of the app is for an incentive-based exercise system that works to measure the effect of physical activity on blood glucose levels, as well as evaluate the feasibility of extrinsic motivation through rewards for patients to engage in regular exercise. The exercise module allows users to set up daily step goals, track completed steps, and monitor calories burned. Detailed insights on steps taken, steps from the previous day, and calories burnt are provided as well as rewarding of the user's activity (Figure 13). By providing positive reinforcement, these incentives motivate users to continue pursuing their health goals.

The system is intended to study the effects of reward-based incentives on user behavior, adherence to exercise routines, and the downstream impacts (long-term) on diabetes management. Such understanding is necessary for maintaining engagement in diabetes self-management. The virtual coin rewards and the goal to hit a certain amount of daily steps intrinsically incentivize the user to stay engaged due to the simple fact of being rewarded. By rewarding coins for daily steps completed, users are getting instant gratification and motivated to continue their exercise. The app's combination of reward system with the ability to track and monitor promotes long-term adherence to diabetes self-management practices.

## 3. Phase III: Proposed Algorithm and Application Evaluation

### 1) Proposed Algorithm Evaluation

The algorithm was evaluated by an expert specializing in the digital health domain. Various scenarios were created, and the corresponding values were entered into the app based on these scenarios. Before the evaluation, each scenario was carefully compared with clinical guidelines to ensure accuracy and relevance. After that, the data was entered into the app to observe the recommendations generated by the algorithm. Each evaluation scenario included at least one decision-making node from the decision process outlined by the algorithm. The expert reviewed each scenario with the appropriate recommendation based on established clinical guidelines.

The data for this evaluation consisted of scenarios commonly encountered in clinical practice, and the recommendations generated by the app were compared to these standards. A scoring system was used to assess the algorithm's accuracy: a score of +1 was assigned when the app's recommendation matched the clinical practice guidelines, a score of 0 when the same recommendation was generated for different scenarios, and a score of -1 when the app's recommendation deviated from the clinical guidelines. The proficiency and efficiency of the algorithm were calculated using the following formulas [51, 52]:

$$Proficiency = \frac{Sum\ of\ (+1, 0\ and\ -1)\ scores}{Total\ number\ of\ Recommendation} \times 100 \qquad Equ.7$$

$$Efficiency = \frac{Sum\ of\ (+1)\ scores}{Total\ number\ of\ Recommendation} \times 100 \qquad Equ.8$$

The proficiency indicates the overall consistency of the app recommendations between the experts and the app to the total number of recommendations. On the other hand, efficiency measures the ratio of consistent recommendations from both the experts and the app to the total number of recommendations. One expert evaluated the proficiency of the algorithm and scored 90% on proficiency and 92% on efficiency. In this evaluation, we examined important areas of the app such as recommending exercises, tracking blood glucose, logging calories and other notifications generated by the app containing reward-based motivation. Subsequently, the scenarios were therefore tailored to represent this particular emphasis on the app's reward and exercise-related features.

### 2) Mobile Application Evaluation (heuristics approach)

A heuristic evaluation is a technique that finds problems in the user interface design to evaluate the usability of computer software. In this procedure, evaluators examine the interface and determine how well it conforms to accepted usability standards. The app's heuristic evaluation was conducted using Bertini's [53] mobile heuristics tool by three experts, two with experience in digital health informatics and one in software development. Eight core principles of mobile heuristics were assessed on a 5-point Likert scale (ranging from 0, indicating 'no heuristic problem,' to 4, meaning 'a significant heuristic problem that must be addressed). The principles evaluated included: visibility of app status, alignment with real-world usage, consistency and mapping, ergonomic and minimalist design, ease of input and screen readability, flexibility and efficiency of use, personalization, aesthetics and privacy, and realistic error management [53, 54]. An Android smartphone was used to install the DSM app and each expert had a day to try out the app and report any heuristic problem they found. If at least two experts reported the same issue, or if a single expert rated the severity of a problem higher than 4, the app was modified accordingly to address those concerns.

Table 4. Heuristic evaluation of prototype and their severity scores

| **Heuristics Problems** | **Severity Scores** | | | **Comments** |
|---|---|---|---|---|
| | **Rater1** | **Rater2** | **Rater3** | |
| Login issues with kakao and Google sign in | 2 | 1 | 2 | Email is not appearing in sign in dialog. |
| Password reset code problem | 1 | 2 | 0 | Password reset code did not received |
| Required scroll bar in home screen | 1 | 1 | 0 | |
| No back button in some screens | 0 | 1 | 1 | |
| Button text visibility issues | 2 | 1 | 1 | |
| Voice buttons accuracy | 3 | 2 | 1 | Accuracy of voice command response |
| Font size, colours and icons issues | 0 | 1 | 0 | The use of font family and size were inconsistent from screen to screen |
| Version compatibly with older phones (below marshmallow) | 0 | 0 | 1 | App is not compatible with some android version |
| No privacy policy to users | 4 | 3 | 4 | No privacy policy is added for user consent |
| Button constrain layouts problems | 2 | 3 | 0 | |

Table 5. App heuristics tools are assessed on 5 -point scale: 0= no heuristics problem, 1=Cosmetic problem only. Need not be fixed, 2=Minor usability. Fixing be given low priority, 3=Major usability problems. Fixing be given high priority, 4=Usability catastrophes. Must be fixed.

The app was heuristically evaluated and 10 issues were found (Table 4). Severity scores and expert feedback were used to address these issues. However, some issues could not be completely fixed, like version compatibility. This happened when devices were running outdated Android OS versions. Two evaluators scored the most critical issue of 4 severity due to its relation with the app's privacy policy and it was prioritized for immediate resolution. Feedback from the heuristic and usability evaluations was applied to the DSM app. These assessments' comments and recommendations were carefully considered and applied to implement changes. All reported issues were documented carefully throughout the evaluation phase and individually handled. After the app has been modified as necessary, it is put through a debugging and compilation process until any remaining issues are identified and resolved. No other error was reported after this. It is reevaluated to see if the app is going to work and be reliable. After a thorough review and analysis, the app was found to be ready to use.

### 3) User Experience (UEQ)

Evaluation methods include learning about users, their tasks and the context in which they work. User experience is evaluated using a variety of methods, including focus groups, questionnaires, card sorting, contextual inquiries, and interviews. Questionnaires are recognized as a low-cost and efficient tool [55]. Questionnaires enable the collection of data from a larger and geographically dispersed sample, in contrast to interviews, which are typically more time-consuming and more suitable for smaller groups [56]. User Experience Questionnaire (UEQ) is an example of a widely used questionnaire for user experience evaluation. We used a 26-item UEQ consisting of six categories of items to assess both the traditional usability aspects (efficiency, perspicuity, dependability) and user experience dimensions (stimulation, novelty) [57].

In this research, the UEQ was employed to evaluate key aspects of the user experience, specifically attractiveness, pragmatic quality, and hedonic quality. Attractiveness quantifies the overall perceived value of the application, while hedonic quality assesses the extent to which the application supports users' needs for development, originality, and stimulation. Pragmatic quality, on the other hand, measures how well the application enables users to achieve their goals effectively. Figure 14 illustrates the specific categories and items evaluated by the UEQ, which formed the basis for the design of the questionnaire in this study.

### 4) Proposed Application Usability Evaluation (UEQ)

The user experience was evaluated with the use of a User Experience Questionnaire (UEQ) [53]. In this study, twenty-five participants were recruited through targeted social media outreach to the digital health and research community. All participants were at least 20 years old, had a master's or PhD degree and had academic or professional background in bioinformatics or digital health. The users were told to install and use the app for a week and fill in a questionnaire through Google Forms. In addition, 10 patients used the app over 21 days. Overall, Thirty-five participants contributed to this evaluation study.

We conduct the user experience assessment with the User Experience Questionnaire (UEQ), which is simple in data analysis and covers a broad range of the dimensions of user experience. The UEQ is a widely used tool for product assessments and it is a reliable metric that covers all the critical dimensions of user experience. It consists of 26 items, each rated on a 7-point Likert scale, and is divided into six dimensions: The variables attractiveness (6 items), efficiency (4 items), dependability (4 items), perspicuity (4 items), novelty (4 items), and stimulation (4 items) (Figure 14). Responses range from -3 (extremely negative) to +3 (extremely positive) with 0 indicating a neutral position.

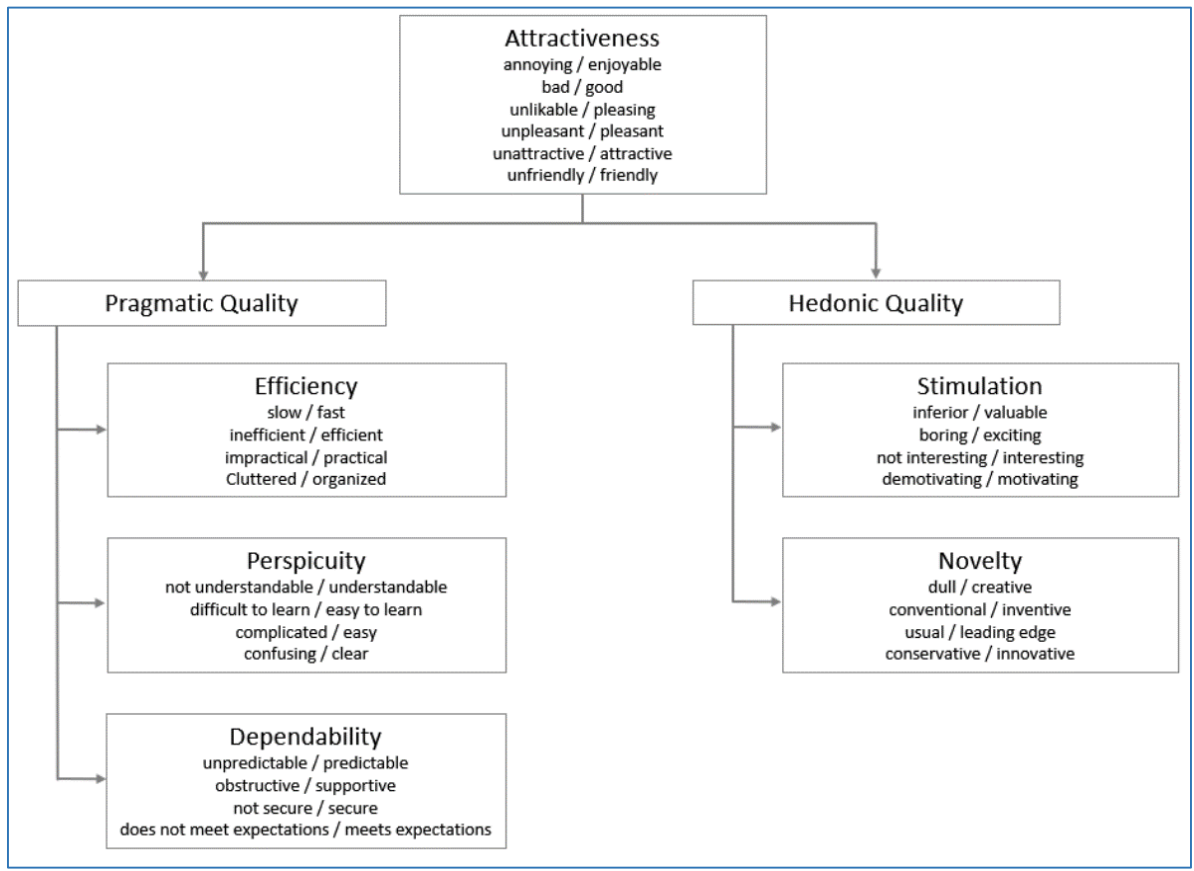

Figure 14. Assumed scale structure of the UEQ.

As seen in Figure 15, every single item value is above the neutrality threshold (0), indicating that users generally had a favorable experience with the diabetes mobile application. With a mean score of 1.2, the item with the lowest rating, "not secure/secure," suggests that users are somewhat concerned about the application's security. Since this is a prototype implementation, the security protocol is not being implemented properly, which results in a low score. Concerns regarding data privacy and protection, which are crucial in applications related to health, may have been raised by the absence of implemented security features.

Conversely, "slow/fast" and "demotivating/motivating," which both received a score of 2.5, are the highest rated items. These findings suggest that users found the application to be not only fast and responsive but also inspiring in terms of motivating them to exercise and manage their diabetes.

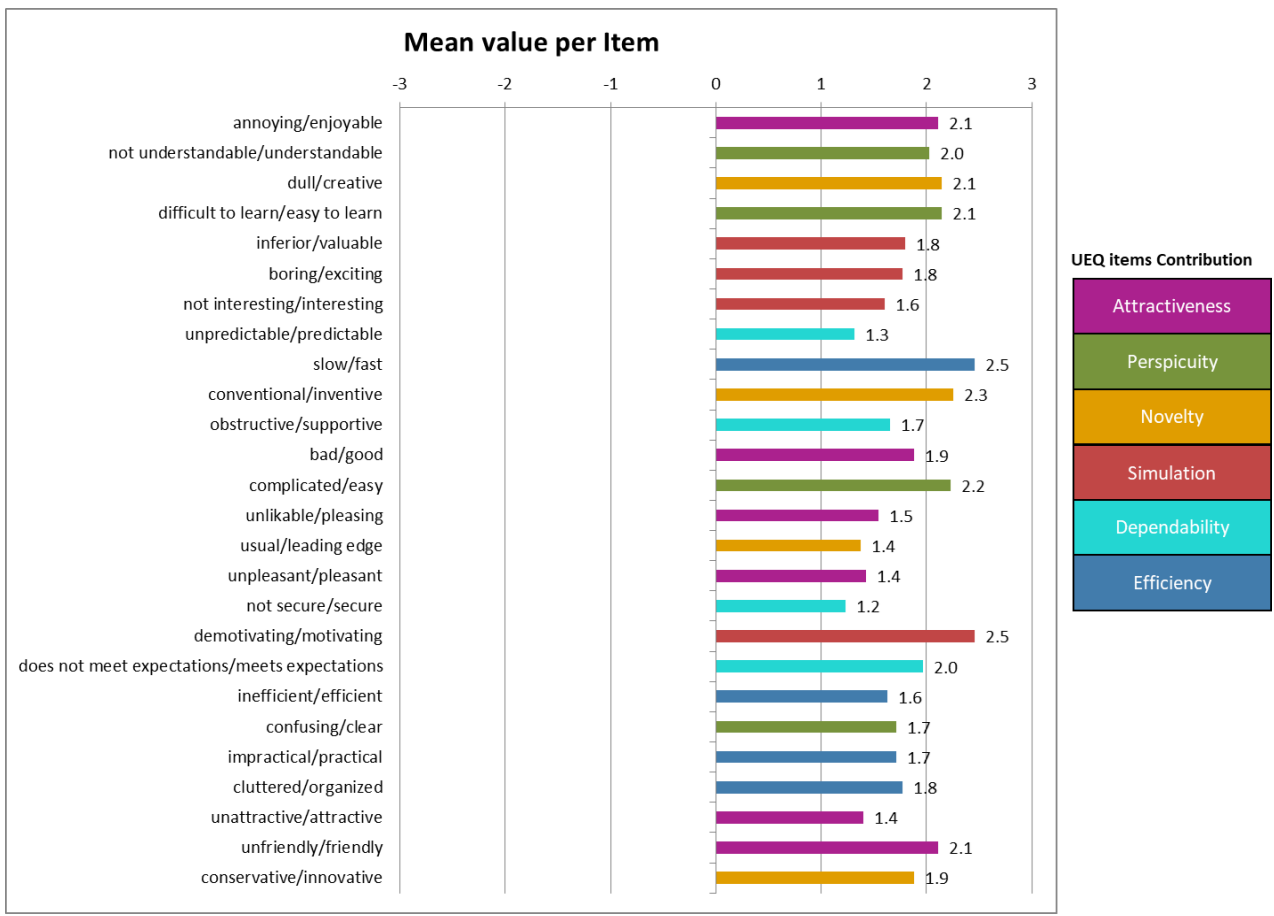

Figure 15. App user experience questionnaire mean value per item

Similarly, in Figure 16, we compute the score of users across six-dimensional scales. In this representation, scores between -0.8 and +0.8 signify a neutral evaluation, values greater than +0.8 indicate a positive assessment, and values less than -0.8 represent a negative evaluation. While the Dependability scale scores lower than the other five dimensions, due to items like "not secure/secure and unpredictable/predictable," the overall evaluation portrays a positive user experience across all six-dimensional scales, with values exceeding 1.5 (Figure 16).

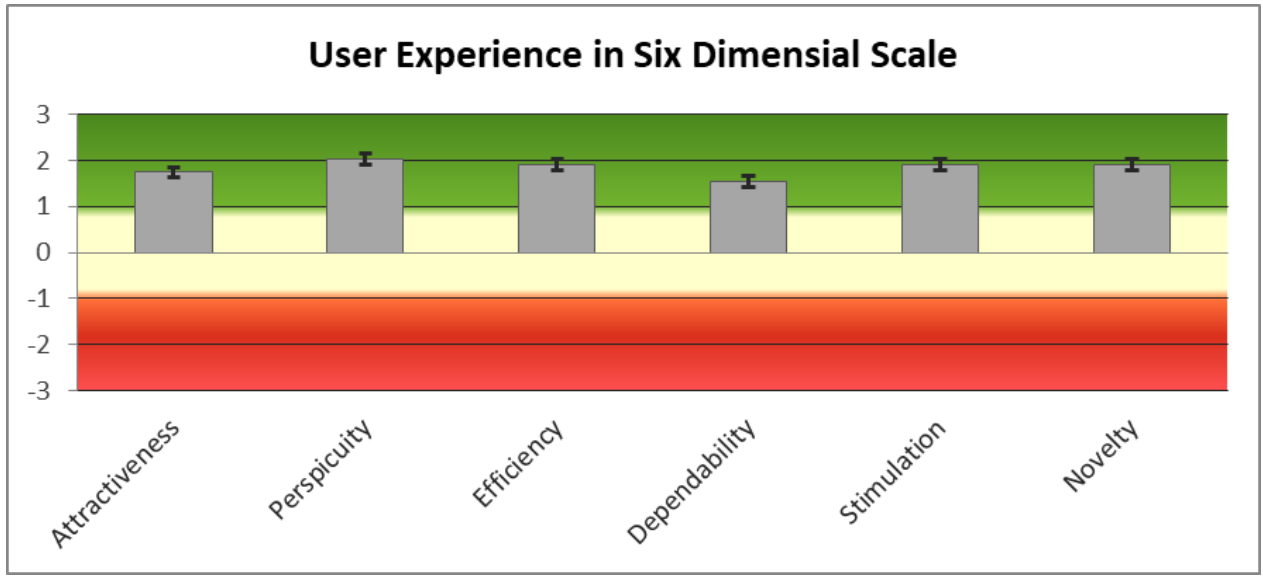

Figure 16.App user experience questionnaire resulting scores on a six-dimensional scale.

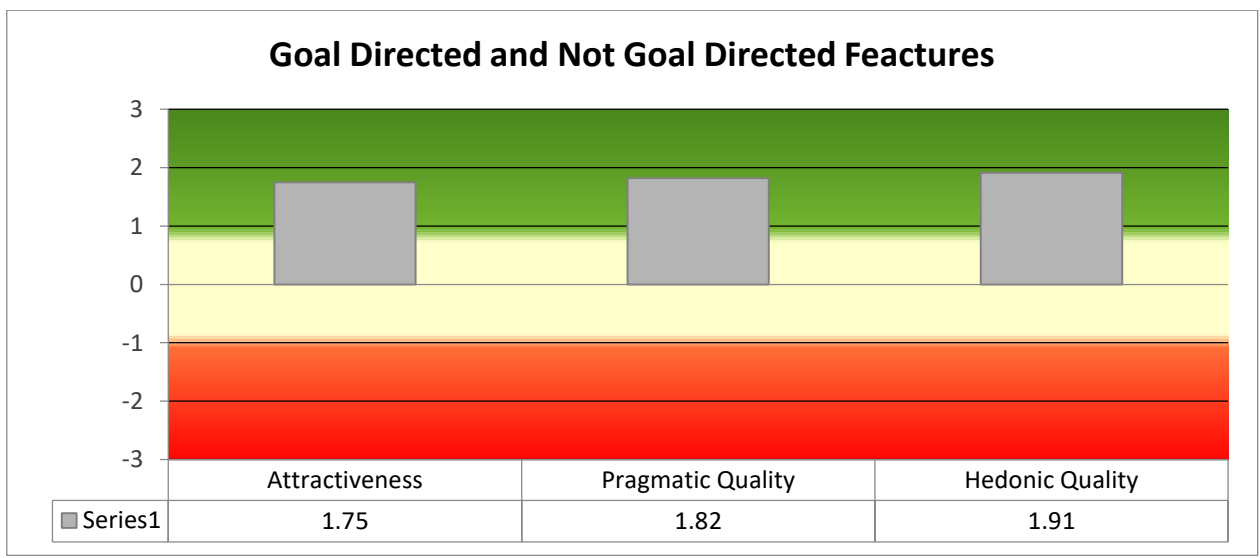

Figure 17.App cumulative score from the user experience questionnaire encompassing the pragmatic and hedonic aspects.

Likewise, the six-dimensional scales are categorized into pragmatic quality (efficiency, perspicuity, and dependability), and hedonic quality (stimulation, novelty) (Figure 17). The pragmatic dimension encompasses quality aspects related to tasks and is oriented towards achieving goals, while the hedonic dimension pertains to quality aspects unrelated to tasks and lacks goal-directedness. In addition, Figure 17 shows that UEQ's attractiveness, pragmatic and hedonic qualities have values of 1.75, 1.82 and 1.91 respectively, reflecting a positive evaluation according to UEQ standards. We also compared the implementation to the UEQ benchmark dataset consisting of 452 product evaluations by 20,190 participants.

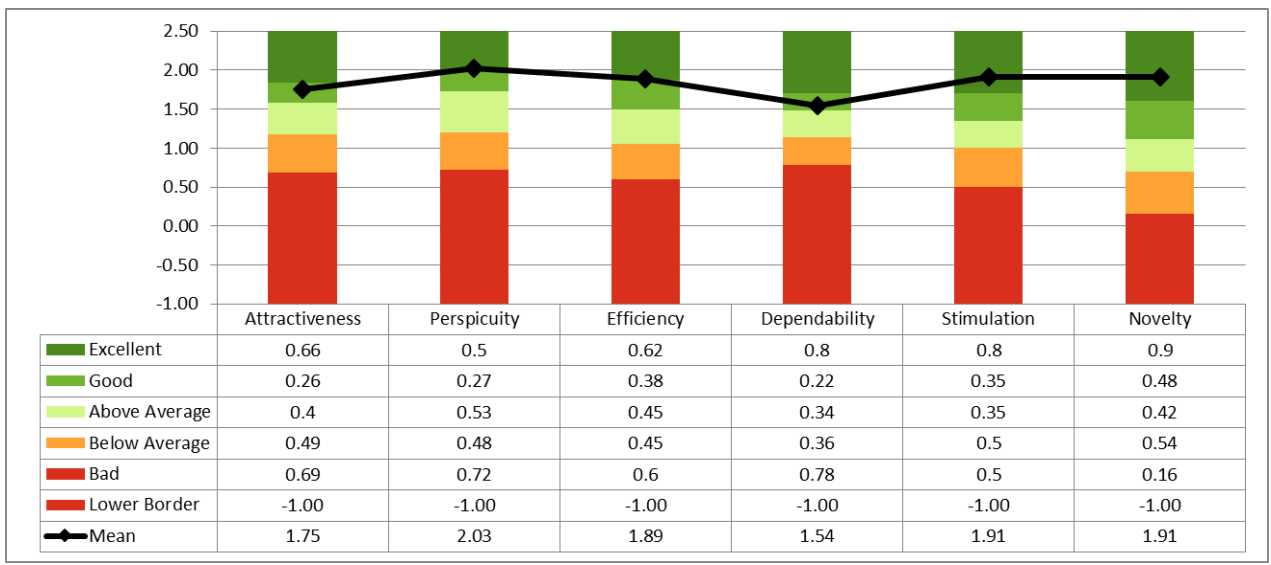

Figure 18. App ratings from the user experience questionnaire across six dimensions align with benchmark data.

The benchmark categorizes a product or system into five classifications for each scale: "Excellent," "Good," "Above Average," "Below Average," and "Bad." The results indicate that the application demonstrates overall effectiveness across all dimensions based on the benchmark data (Figure 18).

# V. Conclusion

The purpose of this study is to investigate a mobile application that uses rewards to encourage people with Type 2 Diabetes to engage in regular exercise, medication adherence, diet control and frequent BG checks to improve their blood glucose control. Our primary aim was to create and assess the development of an app that facilitates the use of incentives to encourage continued participation in healthy behaviors, yet utilizes personalized exercise recommendations and health goals.

The positive correlation between exercise and blood glucose outcomes was clear. The approach was very successful in encouraging users to engage in achieving health goals. Our algorithm performed well with Proficiency and Efficiency scores of 90% and 92%, respectively. Additionally, the app scored well in terms of user experience and was well-received in terms of attractiveness, hedonic quality and pragmatic functionality.

The research confirms that a diabetes management app that includes incentive-type interventions does indeed motivate users to exercise regularly and achieve health goals. Not only does this app help users cope with Type 2 Diabetes, but it also helps users become more physically active and has a potential long-term positive effect on blood glucose control. However because of factors like diabetes severity, the types of foods consumed and exercise routines, responses to exercise vary among individuals.

The results presented in this research show the importance of a simple but effective reward system of virtual coins or points, and that immediate, tangible incentives can have a huge impact on exercise goals and blood glucose management. Gamification helps mHealth by fostering engagement and promoting sustained health goals commitment for the management of chronic diseases in a useful, tech-driven manner. Furthermore, this demonstrates the advantages of a more individualized, incentive-based strategy for enhancing metabolic health.

According to the study, adding more incentives along with appropriate instructions can help users enjoy exercising and self-management. At the same time, it suggests gamification as it positively influences the user's health. To improve exercise routine enjoyment, future research should look into implementing a variety of incentives and user-specific guidelines. If we are adding more advanced gamification features to the app, the app will become more engaging and interactive which will help to improve user health outcomes further.

This study shows that with enough motivation, users can learn to practice healthier routines that can directly benefit their diabetes management. Such a reward-based mHealth strategy could generate a socially innovative impact in taking a proactive way in achieving long-term health benefits.

1. **Limitation and Future Work**

This study has certain limitations that could make it difficult to interpret the results. First, the results' limited generalizability due to the small sample size of 10 participants makes it difficult to extrapolate the intervention's effects to a larger population with Type 2 Diabetes. Additionally, the study's 21 days may not capture the long-term effects of incentive-based exercise on blood glucose control, as extended timeframes could provide more comprehensive insights into sustained behavior changes. Another limitation is the challenge of constant patient monitoring, as real-life variables and external factors may have influenced participant engagement and compliance with the app.

Future studies should try to overcome these constraints by increasing the study's duration and sample size to generate more reliable data and broadly applicable findings. Engaging with a more diverse group of participants over a longer period might give a clearer understanding of the long-term efficacy in terms of incentive-based interventions. Furthermore, clinicians (such as doctors and nurses), who are broadly involved in the study, could provide a controlled environment and assist with regular clinical assessments. In-depth interviews, scheduled follow-up meetings, and patient histories could also be used to augment data for larger samples in order to gather more information. Other relevant factors to be further studied include individual differences in diabetes severity, lifestyles and preferences for personalization and efficacy of future app based interventions.

# 2형 당뇨병 환자의 혈당 조절을 위한 모바일 기반 인센티브 기반운동

와심 아바스

강원대학교 대학원 전자정보통신공학과

초록

모바일 애플리케이션과 스마트폰의 사용이 빠르게 증가하고 있으며 모바일 건강(mHealth) 솔루션은 2형 당뇨병과 같은 만성 질환의 자가 관리 및 교육을 위한 리소스를 제공하는 비용 효율적인 방법을 제공합니다. 당뇨병 관리를 위해 많은 애플리케이션이 개발되었지만, 종종 설계 원칙과 견고한 이론적 기초가 부족하거나 특히 디지털 문해력이 낮은 노인에게 맞춰져 있습니다. 이 연구는 보상 기반 운동 및 당뇨병 자가 관리 애플리케이션을 개발하여 이러한 격차를 해소하는 것을 목표로 합니다. 이 애플리케이션은 사용자가 당뇨병 관련 지표를 편리하게 모니터링할 수 있도록 하고 규칙적인 운동, 약물, 식단 및 빈번한 BG 검사를 통해 상태를 관리하도록 권장합니다.

　우리는 당뇨병 환자의 라이프스타일을 개선하는 것을 목표로 하는 인센티브 기반 추천 알고리즘을 제안하고 개발합니다. 이 알고리즘은 개인화된 건강 추천을 제공하기 위해 실제 모바일 애플리케이션에 통합됩니다. 처음에 사용자는 걸음 수, 칼로리 섭취량, 성별, 나이, 체중, 키 및 혈당 수치와 같은 데이터를 입력합니다. 데이터가 사전 처리되면 앱은 개인화된 건강 및 포도당 관리 목표를 식별합니다. 추천 엔진은 이러한 목표에 따라 운동 루틴과 식단 조정을 제안합니다. 사용자가 목표를 달성하고 이러한 추천을 따르면 인센티브를 받아 준수를 장려하고 긍정적인 건강 결과를 촉진합니다. 또한 모바일 애플리케이션을 통해 사용자는 설명적 분석을 통해 진행 상황을 모니터링할 수 있으며, 이를 통해 일상 활동과 건강 지표가 그래픽 형태로 표시됩니다.

　제안된 방법론을 평가하기 위해, 2형 당뇨병을 앓고 있는 10명의 참가자를 대상으로 3주 동안 연구를 수행했습니다. 참가자는 광고와 건강 전문가의 추천을 통해 모집되었습니다.

환자의 휴대전화에 애플리케이션을 설치하여 3주 동안 사용했습니다. 전문가는 또한 환자의 건강 기록을 모니터링하여 이 연구에 참여했습니다. 알고리즘의 성능을 평가하기 위해 효율성과 능숙도를 계산했습니다. 그 결과, 알고리즘은 각각 90%와 92%의 능숙도와 효율성 점수를 보였습니다. 마찬가지로, 우리는 매력성, 쾌락적, 실용적 품질 측면에서 애플리케이션에 대한 사용자 경험을 계산했으며, 연구에 참여한 35명을 대상으로 했습니다. 그 결과, 전반적으로 긍정적인 사용자 반응을 보였습니다. 연구 결과에 따르면 운동과 보상 사이에 명확한 긍정적 상관관계가 있으며, 운동 후 사용자 결과에서 눈에 띄는 개선이 관찰되었습니다. 인센티브 기반 시스템은 사용자가 규칙적인 신체 활동에 참여하도록 효과적으로 동기를 부여하여 당뇨병 자가 관리 결과를 개선합니다.

□ 핵심 단어

운동, 2형 당뇨병, 보상(게임화), 혈당 관리, 모바일 건강(mHealth), 신체 활동, 동기 부여 및 행동 변화.